# Realization of an inherent time crystal in a dissipative many-body system

Yu-Hui Chen[1, 2] and Xiangdong Zhang[1, 2, *]

[1] Key Laboratory of Advanced Optoelectronic Quantum Architecture and Measurements of Ministry of Education, School of Physics, Beijing Institute of Technology, 100081, Beijing, China

[2] Beijing Key Laboratory of Nanophotonics & Ultrafine Optoelectronic Systems, School of Physics, Beijing Institute of Technology, 100081, Beijing, China

* Corresponding author: zhangxd@bit.edu.cn

**Time crystals are many-body states that spontaneously break translation symmetry in time the way that ordinary crystals do in space. While experimental observations have confirmed the existence of discrete or continuous time crystals, these realizations have relied on the utilization of periodic forces or effective modulation through cavity feedback. The original proposal for time crystals is that they would represent self-sustained motions without any external periodicity, but realizing such purely self-generated behavior has not yet been achieved. Here, we provide theoretical and experimental evidence that many-body interactions can give rise to an inherent time crystalline phase. Following a calculation that shows an ensemble of pumped four-level atoms can spontaneously break continuous time translation symmetry, we observe periodic motions in an erbium-doped solid. The inherent time crystal produced by our experiment is self-protected by many-body interactions and has a measured coherence time beyond that of individual erbium ions.**

**Introduction:**

Similar to ordinary crystals where atoms take periodic positions in space, time crystals are many-body states that spontaneously recur in time. However, a system spontaneously repeating its pattern implies the breaking of time translation invariance, which contradicts the time-independence of most states in conventional theory. The realization of time crystals was first addressed by Wilczek[1,2], but subsequent no-go theorems indicated that they could not exist in thermal equilibrium states[3--5]. Nevertheless, recent advancements have led to the realization of driven discrete time crystals[6-8] in close quantum systems, characterized by oscillations at twice the driving periods[9-17]. Yet, the heating associated with driving in these closed systems prevents the persistence of time crystal. Theoretical research suggests that dissipation may overcome the heating issues [18-24], leading to the observations of dissipative discrete time crystals[25,26]. Moreover, if the driving becomes non-periodic and time-invariant, the studied system acquires continuous time translation symmetry. Although the potential heating problem of a continuous driving can be worse compared to that of a periodic force, continuous time translation symmetry can also be spontaneously broken in dissipative systems[18-24,27]. Recent experimental observation has confirmed the existence of continuous time crystal in an atom-cavity system[28]. Like periodically driven systems, there remains a built-in frequency stemming from the optical cavity[22,23], which introduces an effective periodic modulation.

The aforementioned discrete and continuous time crystals depend on the imposition of external periodic constrains, such as periodic forces or cavities, to break time translation symmetry. However, according to the spirit of the original proposal, time crystals represent the spontaneous emergence of time-periodic motions within time-invariant systems. These motions are inherently self-triggered and self-sustained without the need of introducing external periodic inputs. However, creating such an inherent phase is still a pending challenge.

Here we report in both theory and experiment that inherent time crystals can be realized in a dissipative quantum system, which represents built-in phases of many-body systems that do not rely on recurring forces or cavities. By exploiting the dipole-dipole interactions within an erbium-ion ensemble, we not only observe the spontaneous breaking of continuous time translation, but also the emergence of temporal order. The resultant self-sustained motion exhibits a period determined by the system's parameters and is protected by many-body interactions from inner degrees of freedom. Furthermore, the persistence of the time crystal's oscillations reveals long-range time correlation beyond the coherence time of individual ions.

**Results:**

**Theoretical Model**

We investigate a system consisting of a collection of four-level atoms that are driven by a continuous-wave (CW) laser, as depicted in Fig. 1a and 1b. There are a pair of Kramers doublets for both the optical ground and excited states. The atoms possess electron spins and interact with one another via magnetic dipole-dipole interactions. The system Hamiltonian is the sum of the Hamiltonian of individual atoms, which is

given by

$$H_{\text{sys}} = \sum_i \left( \sum_{g,e} \delta_{g,i} |g\rangle\langle g|_i + \Delta_{e,i} |e\rangle\langle e|_i + \Omega_{ge}(\mathbf{r}_i)|g\rangle\langle e|_i + \Omega_{ge}(\mathbf{r}_i)|e\rangle\langle g|_i \right) \\ + \sum_{i,j} \frac{J_{ij}}{|\mathbf{r}_{ij}|^3} [\mathbf{S}_i \cdot \mathbf{S}_j - 3(\mathbf{S}_i \cdot \hat{\mathbf{r}}_{ij})(\mathbf{S}_j \cdot \hat{\mathbf{r}}_{ij})], \quad (1)$$

where $|g\rangle_i$ and $|e\rangle_i$ indicate the optical ground and excited states of the $i$th atom, the marks below summation $g (= 1 \text{ or } 2)$ and $e (= 1 \text{ or } 2)$ describe two doublets in the ground and excited states, $\delta_g$ and $\Delta_e$ are the frequencies of the ground and excited states, $\Omega_{ge}(\mathbf{r}_i)$ is the Rabi frequency, $J_{ij}$ is the strength of the dipole interactions, $\mathbf{S}_i$ is the magnetic dipole moment of the $i$th atom, $\mathbf{r}_{ij}$ is the vector connecting two spins $\mathbf{S}_i$ and $\mathbf{S}_j$, and $\hat{\mathbf{r}}_{ij}$ is the corresponding unit vector. The first term on the right-hand side of Eq. (1) represents the optical Bloch description of the light-matter interactions. The second accounts for the many-body interactions between the $i$th and the $j$th atom, which can lead to an excitation-dependent frequency shift to the erbium ions [See Supplementary Note 1].

Hamiltonian shown by Eq. (1) can describe various quantum systems[29-32], and is particularly relevant to erbium-doped crystals. In these crystals, the erbium ions exhibit an optical transition of 1.5 μm, and the effective spins for both their optical ground and excited states are $S = 1/2$. Due to the presence of an average magnetic field produced by all other erbium ions in the crystal, the spin states experience a slight splitting. As a result, the erbium ions can be considered as a nearly-degenerate four-level system, as shown in Fig. 1b. When there is no light to drive the erbium ions, all ions remain in their ground states, and the optical resonant frequencies of neighboring ions are of no difference (as shown at the top of Fig. 1a). However, this situation changes if some ions are optically excited[33]. Generally, the magnetic dipole moments of ions in their optical excited states differ from that in the ground. Therefore, optically exciting an ion can instantaneously change the local magnetic field seen by its neighboring ions. The optical frequencies of nearby ions are thus modified as a result of the Zeeman effect[34,35], as illustrated at the bottom of Fig. 1a, which in turn affects the absorption of light. Moreover, the presence of optical spontaneous emission, spin relaxation, and decoherence makes the erbium ions inherently an open quantum system. The combined effect of the openness of the quantum system and the aforementioned excitation-induced effect can give rise to rich non-equilibrium dynamical phases[27].

To explore the physical properties of the four-level erbium system, we use the Runge-Kutta method to calculate the time response of a collection of erbium ions governed by the Hamiltonian Eq. (1). Specifically, by applying the mean-field approximation[36] and introducing the decay and dephasing terms of the ions, we calculate the density matrix $\rho(t)$ in the framework of Lindbald master equation $\dot{\rho} = L\rho$, where $L$ is the Lindblad superoperator. The calculation process is detailed in Method and Supplementary Note 1, and the parameters used are typical values of erbium ions at a temperature of 4 K. The calculated results for the population at the $n$th level $\rho_{nn}(t)$, i.e., the diagonal terms of $\rho(t)$ are shown in Fig. 1c.

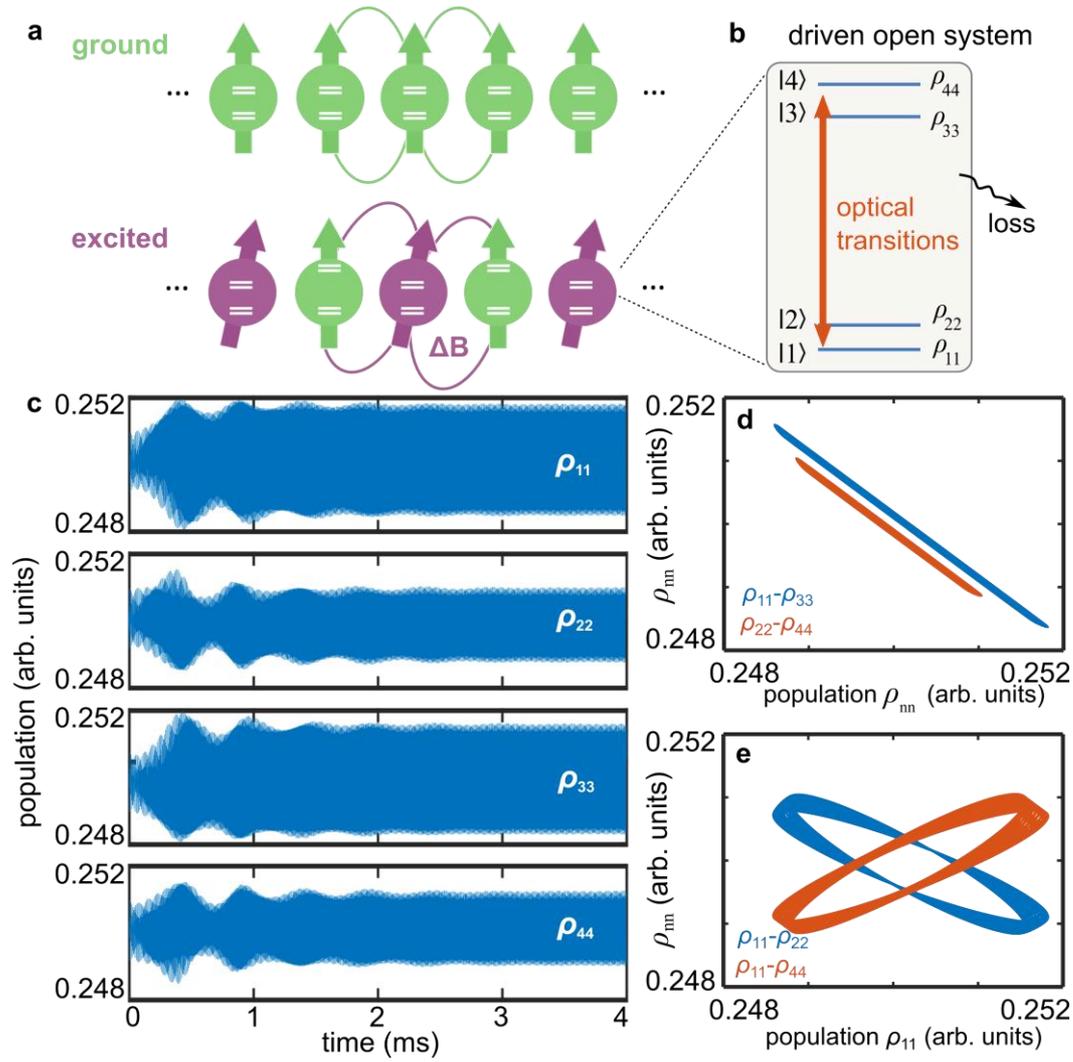

**Figure 1. Theoretical model for realizing inherent time crystal. a.** Magnetic many-body interactions. Atoms in their optical ground and excited states have different magnetic dipole moments. When atoms are optically excited, as shown by the purple spins, they introduce local magnetic field variance $\Delta B$ and thus change the optical resonant frequencies of nearby atoms, as illustrated by the energy levels of the two green spins at the bottom. **b.** The energy structure of our four-level systems. The coupling to their environment leads to the decay and decoherence of the atoms. **c.** The dynamic behaviors of the populations of the four levels. The $\rho_{nn}(t)$ oscillate according to their Rabi frequencies, resulting in a high density of lines in the plot. The calculation begins with an initial state of an equally-populated mixed-state, that is, $\rho_{nn}(0) = 0.25$, to better demonstrate the long-time behaviors of $\rho_{nn}(t)$. **d.** Blue, the relationship between $\rho_{11}$ and $\rho_{33}$; orange, the relationship between $\rho_{22}$ and $\rho_{44}$. **e.** Blue, the relationship between $\rho_{11}$ and $\rho_{22}$; orange, the relationship between $\rho_{11}$ and $\rho_{44}$. The parameters of the calculation can be found in Supplementary Note 1.

It is shown clearly that the populations of the four-level system ($\rho_{11}$, $\rho_{22}$, $\rho_{33}$ and $\rho_{44}$) oscillate persistently instead of relaxing to a stationary state. In conventional systems driven by time-invariant forces, it is widely believed that the system output is also time-invariant in the long-time limit. In principle, the emergence of any oscillations with frequency $\omega \neq 0$ indicates that the time translation symmetry is violated. For our calculated system, since it is governed by a time-invariant Hamiltonian Eq. (1), the corresponding outputs have to be, if the time translation symmetry holds, time-independent in the long-time limit. However, if the time translation symmetry is spontaneously broken, then the relaxation to stationarity is no longer guaranteed. The emergence of the persistent oscillations in Fig. 1c strongly suggests a breaking of the time symmetry in our system. Moreover, a Fourier analysis of the data $\rho_{nn}(t)$ in the long-time limit shows a peak at the frequency of 46.4 kHz (see Note 2 in SI for more details), further indicating the formation of temporal order.

To provide further clarification on the physics of the persistent oscillation, we examine the relations between different $\rho_{nn}(t)$ in Fig. 1d and 1e. While $\rho_{11}$ and $\rho_{22}$ are nearly-linear functions of $\rho_{33}$ and $\rho_{44}$, respectively, the dependences of $\rho_{22}$ and $\rho_{44}$ on $\rho_{11}$ show the behaviors of limit cycles[37], as shown in Fig. 1d and 1e. In our calculation, the ratio of the coupling strength of the transition $|1\rangle - |3\rangle$ to the transition $|1\rangle - |4\rangle$ is 1.87:1 in the calculation. Thus, the optical field is more likely to drive the transition between $|1\rangle$ and $|3\rangle$ rather than that between $|1\rangle$ and $|4\rangle$. Consequently, the population difference of $\rho_{11}$ and $\rho_{33}$ is primarily governed by the balance of optical pumping and dissipations, resulting in the synchronization of $\rho_{11}$ and $\rho_{33}$ through the optical Rabi oscillation, as shown in Fig 1d (a similar conclusion also holds for the relations of $\rho_{22}$ to $\rho_{44}$ and $\rho_{22}$ to $\rho_{33}$). In contrast, the presence of limit cycles between $\rho_{22}$ and $\rho_{11}$ in Fig. 1e suggests that the changes of $\rho_{11}$ and $\rho_{22}$ are not synchronized under the same optical driving. Instead they undergo a competing process.

It is the competition between different optical transitions that prevent the system from reaching a stationary state. For a pure two-level system, where there is only one possible optical transition, it is impossible to generate a competing process by inner degrees of freedom[18,27]. To observe an inherent time crystalline phase, the energy structure of interacting atoms needs to have more than two levels. This increased complexity allows the many-body interactions to act as intrinsic nonlinear interactions and provide positive feedback to the competition between different optical transitions (see Supplementary Note 2 and 3 for more details). The interplay between these complex processes leads to the formation of a temporal order. Unlike the time crystals so far demonstrated, the persistent oscillation of $\rho_{nn}$ (Fig. 1c) in such a CW-driven four-level system is purely self-organized, and its recurring frequency is only determined by the coupling parameters of the system itself, suggesting that the time crystalline order is inherent.

**Time crystalline order**

In the following, we discuss how to realize such an inherent time crystal in experiments. We used erbium-doped yttrium orthosilicate (Er:Y$_2$SiO$_5$) with a concentration of 1000 ppm, corresponding to an average distance between erbium ions of approximately 4 nm, such that the magnetic dipole-dipole interactions between erbium ions are on the order of 10 MHz (see Supplementary Note 4 for more details). Additionally, there are two pairs of Kramer doublets separated by an optical resonance of 1536.4 nm (one of the two 1.5 μm transitions of erbium ions in $Y_2SiO_5$ [38]). The effective spins for both their optical ground and excited states are $S = 1/2$. If no magnetic field is applied, the optical ground and excited states of erbium ions are degenerate two levels. However, the spin levels may experience a minor degree of splitting due to the influence of the average magnetic field generated by all the other erbium ions in the crystal. That is to say, the effective spin $S = 1/2$ is split into two levels with small frequency differences $\mathcal{O}(0.1\,\text{MHz})$[27], which makes erbium ions a four-level system as shown in Fig. 1b. If we further apply a CW laser to pump the erbium ensemble to their excited states, then we can construct a system described by Hamiltonian Eq. (1), where the excited ions can change the resonant frequency of their neighbors as shown in Fig. 1a. Under certain conditions, the combined effect of the many-body interactions, the driving field and the losses of erbium ions can prevent the system from reaching a stationary state[27]. A periodic motion can emerge as a result of the inherent competitions between different optical transitions.

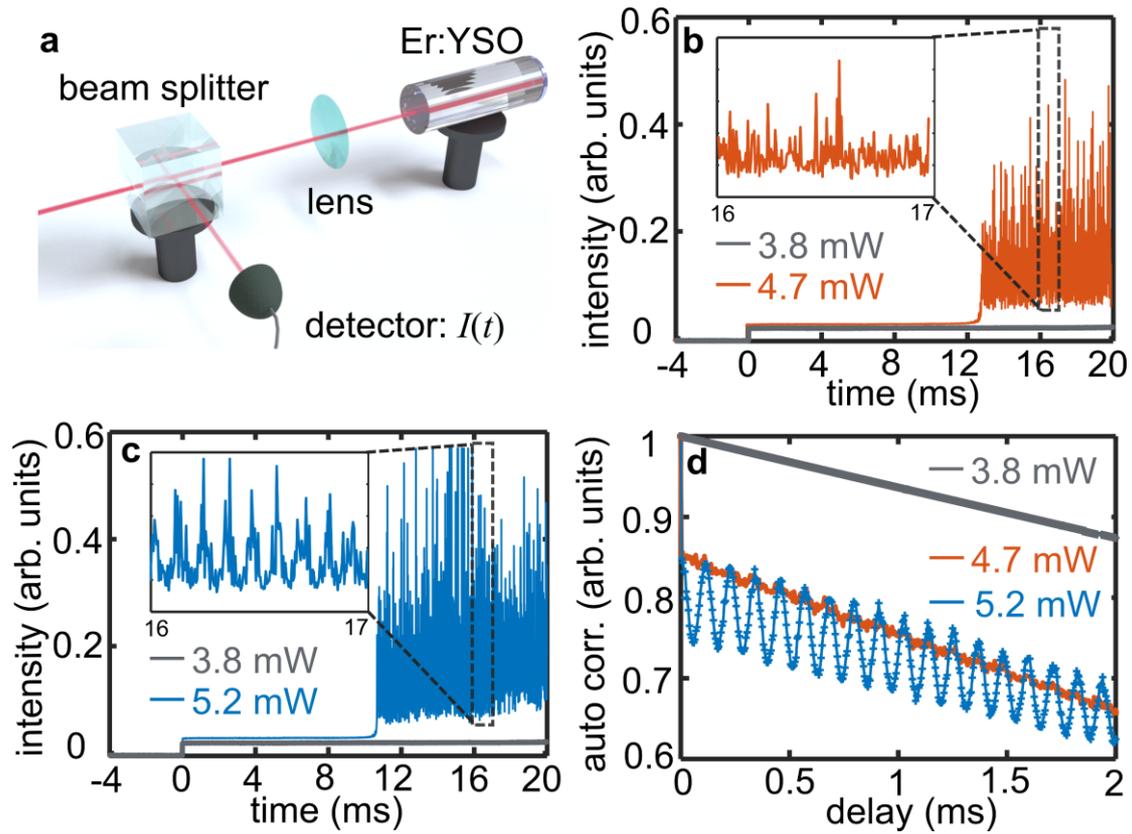

**Figure 2. Inherent time crystal in an erbium ensemble. a.** Schematic drawing of the experimental setup. A laser beam is input from the left side. There is a mirror at the right end side of the Er:Y$_2$SiO$_5$ sample, such that the light double-passes the sample before it hits a detector. The detector gives a time response of $I(t)$. **b.** Experimental realization of broken time translation symmetry without time crystalline order. The laser is switched on at $t = 0$. Gray, the measured $I(t)$ for $P_{in} = 3.8$ mW; blue, the $I(t)$ for $P_{in} = 4.8$ mW. Inset, the $I(t)$ from 16 to 18 ms for $P_{in} = 3.8$ mW. **c.** Experimental realization of broken time translation symmetry together with time crystalline order. Gray, the measured $I(t)$ for $P_{in} = 3.8$ mW; orange, the $I(t)$ for $P_{in} = 5.2$ mW. Inset, the $I(t)$ from 16 to 18 ms for $P_{in} = 5.2$ mW. **d.** Auto-correlation of the data $I(t)$ in (b) and (c) with pump power $P_{in}$ as noted. The laser frequency is set to $f_l = 0.00$ GHz

To investigate the presence of a time crystalline order, we optically excite the erbium ensemble with a CW laser of different frequency $f_l$ and power $P_{in}$. For notational convenience, all quoted $f_l$ in this paper are the frequencies minus

195117.17 GHz (the center frequency of the inhomogeneous line of Er:Y$_2$SiO$_5$). The sample is placed in a cryostat and cooled down to a temperature of 4 K. The light that double-passes the sample is detected as a function of time $I(t)$, as illustrated in Fig. 2a. The measured $I(t)$ is determined by the sample absorption and thus is directly related to $\rho(t)$ (see Supplementary Note 1 and the literature[27] for more details). If there is any oscillation in $\rho(t)$ as theoretically predicted above, $I(t)$ can be utilized to monitor the changes.

Figure 2b and 2c show the measured $I(t)$ as a function of time for two different $P_{in}$, where the laser frequency $f_l = 0.00$ GHz is on resonance with the absorption of erbium ions and the laser is switched on at $t = 0$ ms. When the pump power is low ($P_{in} = 3.8$ mW), the erbium ensemble absorbs most of the laser energy and gives a low output in the detector, as shown by the gray line in Fig. 2b and 2c. The $I(t)$ under this circumstance eventually relaxes to a stationary value, as expected for systems with time translation symmetry. However, if the pump power is increased to $P_{in} = 4.7$ mW, the output $I(t)$ at the beginning is likewise low, but becomes dynamically unstable after a delay of approximately 12 ms, as shown by the blue line in Fig. 2b. Even in the long-time limit $t \to \infty$, such oscillating $I(t)$ persists and has a broad-band frequency response, with a cutoff frequency ~50 MHz[27]. In our system, the driving field is a time-independent CW laser, and the resulting Hamiltonian is time invariant as well. The emergence of an oscillating $I(t)$ in Fig.2b, in contrast to a time-independent output predicted by conventional theory, clearly suggests a non-stationary $\rho(t)$ and evidences that the time translation symmetry is spontaneously broken in the erbium ensemble (see the literature[27] and Supplementary Note 5 for the persistent oscillations).

If we further increase $P_{in} = 5.2$ mW, similar oscillating $I(t)$ can be observed, as shown in Fig, 2c. More importantly, when a segment of $I(t)$ in Fig. 2c is zoomed in, a hidden periodic order can be identified, which in contrast is absent in the magnified view of Fig. 2b. This suggests that a temporally recurring motion, i.e., a time crystalline order, arises if $P_{in}$ is increased from 4.7 to 5.2 mW. To explicitly determine the time crystalline order of our sample, the auto-correlation function $\langle I(t)I(t-\tau)\rangle$ corresponding to Fig. 2b and 2c are presented in Fig. 2d. When the input power ($P_{in} = 3.8$ mW) of the pump is low, the corresponding auto-correlation curve shows linearity, indicating the absence of periodicity in $I(t)$. Upon increasing the input power to 4.7 mW, although the continuous time translation symmetry is broken (as shown in Fig. 2b), the correlation curve remains linear (the orange curve in Fig. 2d). However, at a pump of $P_{in} = 5.2$ mW, a distinct periodicity can be identified, as depicted by the blue curve in Fig. 2d.

The periodic oscillation in $I(t)$ is significantly different from the well-known self-pulsing effect in erbium doped fiber lasers[39-43]. Self-pulsing is due to the dynamic interplay or the competition between the gain and the losses within a laser cavity. As a result, the net gain inside the cavity is periodically modulated and generate fluctuations in the laser output. However, our experiment did not adopt such a cavity configuration as one side of the sample was coated with an anti-reflective layer of less than 0.8 % reflectivity. Consequently, our observations could not have been attributed to the self-pulsing effect in lasers. As aforementioned, the measured $I(t)$ is closely related to

$\rho(t)$. Thus, the periodicity in $I(t)$ under a $P_{in} = 5.2$ mW optical driving indicates a repeating temporal order of $\rho(t)$, which agrees well with the characteristics of a time crystalline order. Together with the spontaneous breaking of time translation symmetry shown in Fig. 2b and 2c, and the lack of imposed periodicity from a driving force or a cavity, the revealed periodicity in Fig. 2c and 2d suggests the self-formation of an inherent time crystalline phase.

**Phase transitions**

The emergence of a time crystalline order as shown in Fig. 2 depends on $P_{in}$ and can be observed for different laser frequencies $f_l$. Shown in Fig. 3a is the measured spectra of $I(t)$ for different pump power (the pump laser is always on during the spectral measurements) and for a different laser detuning $f_l = 0.50$ GHz (see Supplementary Note 6 for more details). When the pump power is low, for example, $P_{in} = 4.0$ mW, no oscillating $I(t)$ can be observed in the long-time limit. The spectral response of such a normal phase is a delta function with its maximum at zero frequency, as shown by the gray curve in Fig. 3a. Increasing the pump power to $P_{in} = 8.0$ mW, $I(t)$ becomes unstable after a delay of approximately 14 ms (see Supplementary Note 7 for more details). This optical instability is verified by the rising nonzero frequency components in the spectrum shown by the yellow curve in Fig. 3a. However, the lack of peak in the spectrum of $P_{in} = 8.0$ mW suggests that periodic temporal order is still absent in $I(t)$. When increasing the pump power to $P_{in} = 9.5$ mW, a sharp peak at 8.7 kHz emerges, as shown by the purple curve in Fig. 3a. The emergence of such a spectral peak depends on the pumping intensity $P_{in}$ and the laser frequency $f_l$, and its frequency is of the same as the correlation curve peak shown in Fig. 2c. This observation suggests that although the varying Rabi frequency due to different $P_{in}$ can indirectly impact the time-crystal formation, the underlying mechanism governing the time crystalline behavior is robust and not directly affected by the Rabi frequency of the driving field. Furthermore, identical time-crystalline frequency is observed when the crystal is rotated by 0, 20, and 60 degrees, even though the crystal is highly anisotropic and the ground state g-factors vary between 2 and 15. This result makes it clear that the oscillating frequency is not caused by interference between two optical transitions split by the Earth's magnetic field or a residual magnetic field. In addition, our calculations indicate that while factors such as decoherence times and lifetimes have some influence on the oscillation frequency, the frequency is mostly sensitive to the competition between different transitions (see Supplementary Note 3).

The curves of $P_{in} = 8.0\,\text{mW}$ and 9.3 mW indicate a transition from a phase with broken time translation symmetry but without temporal order to a phase of self-organized time crystalline order. However, when the pump power was increased to 13.0 mW, the width of the 8.7 kHz peak broadened significantly, as depicted by the blue curve in Fig. 3a. Such a behavior in the spectrum can be attributed to high-order effects induced by a strong optical pump, where the time crystal undergoes oscillations with a broad band of other frequencies and the 8.7 kHz temporal order becomes less distinct. This can cause an imprecise determination of the frequency position. To observe a well-defined temporal order, that is, a time crystal with a long-range temporal correlation or a long coherence time, it is critical to set a proper optical driving field.

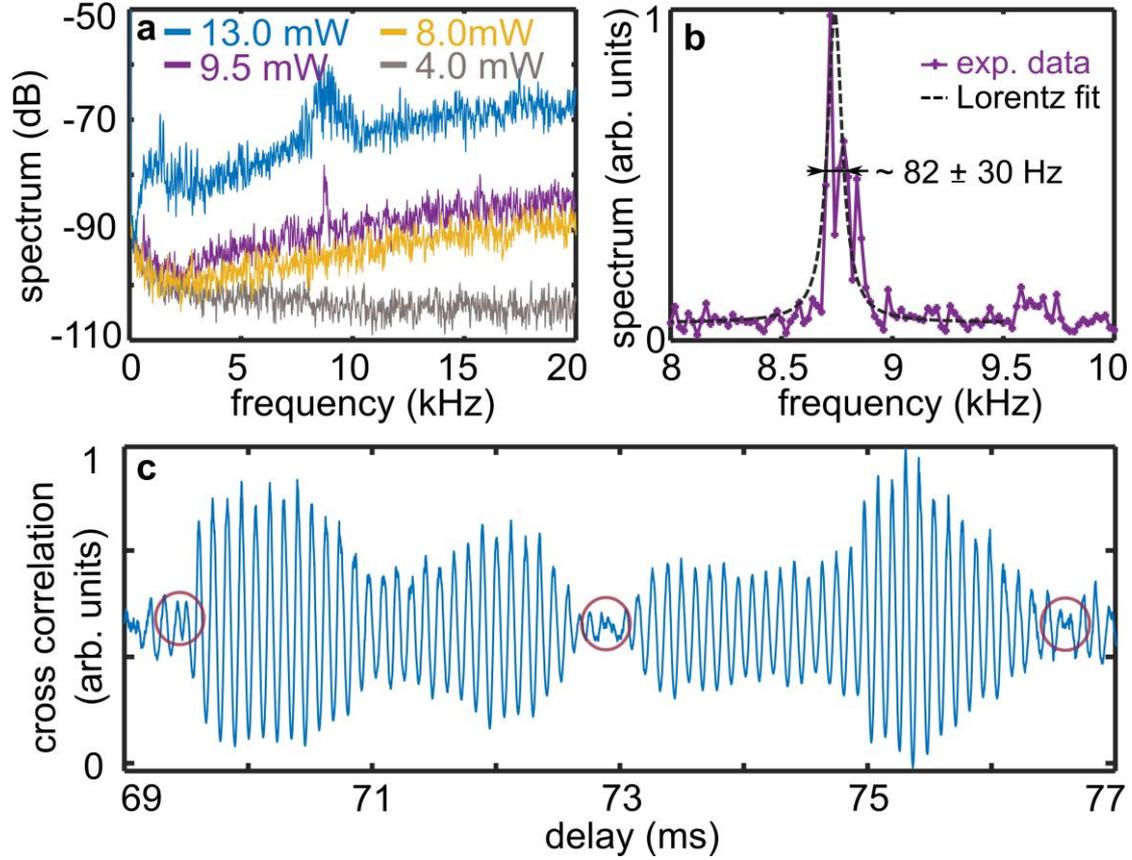

**Figure 3. Phase transitions and coherence time. a.** Phase transitions with pump-power dependence. Measured spectra of the time crystal at different pump power as noted. The laser frequency is $f_l = 0.50$ GHz. **b.** Lorentz fit of the 8.7 kHz spectral peak of $P_{in} = 9.5$ mW. **c.** Phase noise of the inherent time crystal. The cross-correlation $F(\tau)$ shows that the periodicity of $I(t)$ remains even for 70 ms. Time points with phase discontinuity are marked by circles.

Noting that the spectrum of $P_{in} = 9.5$ mW in Fig. 3a has the narrowest linewidth, we use it to determine the coherence time of the time crystal. The full width at half-maximum given by a Lorentz fitting is $\Gamma = 82 \pm 30$ Hz (Fig. 3a), corresponding to a coherence time of $T_2 = 1/\pi\Gamma = 4 \pm 2$ ms, which is beyond that of individual erbium ions (for reference, the coherence time of erbium ions in a 50 ppm Er:Y$_2$SiO$_5$ at 4.2 K is on the order of 10 $\mu$s[44]). Ideally, periodic oscillations of a time crystal maintain phase coherence over indefinitely long periods of time[45,46]. Because the long-range temporal order of a time crystal is protected by many-body interactions and its coherence is robust to the dephasing effect from the environment.

Unlike an ideal one, our time crystal shows a finite coherence time in the long-time limit (Fig. 3b), suggesting that its periodic oscillations are consecutive and dephase in a time scale of ~ 4 ms. To reveal the dephasing on the oscillations of $I(t)$, we define a cross-correlation function $F(\tau) = \langle I(t)r(t-\tau)\rangle$, where $r(t)$ is a reference pulse function oscillating with 8.7 kHz. Such that $F(\tau)$ reproduces the phase information of $I(t)$ for $\tau = t$ (see Method and Supplementary Note 8 for detailed discussions).

As an example, the consecutive periodicities of our time crystal for the delay of 69 - 77 ms are shown in Fig. 3c [$F(\tau)$ at other delays can be found in Supplementary Note 9]. It is clear that the oscillations shown in Fig. 3c can be divided into segments with temporal lengths of ~ 4 ms, among which phase discontinuities arise, as marked by the circles in Fig. 3c. These phase discontinuities indicate that random phase shifts arise at the corresponding time points, which interrupt $I(t)$ from being a perfectly periodic signal. That is to say, the present result is subjected to modulations from phase noise on a time scale of ms.

Note that the root-mean-square stability of the intensity of our laser is 0.2% in ms scale, and its frequency drift, caused by temperature fluctuations, is approximately 0.2 MHz/s. These parameters are not significant enough to produce any noticeable effects on our system. However, our driving laser has a phase-noise-limit linewidth of less than 100 Hz. We then performed a calculation to estimate how phase shifts in the driving field affect the inherent time crystal (see Supplementary Note 10 for more details). The results show that phase perturbations in the driving can break the time crystalline oscillations into consecutive segments similar to what is shown in Fig. 3c. Given that the linewidths of our time crystal and that of our laser are in the same order, together with the similar phase-modulated behaviors of our calculated results (Supplementary Note 10) and the experimental results (Fig. 3c), we ascribe the dephasing of our time crystal to the phase noise of the driving laser.

Such a dephasing time, approximately 4 ms, is significantly longer than the coherence time of individual erbium ions in the YSO crystal[44], which suggests the protection of the crystalline order by many-body interactions beyond the dephasing nature of individual ions. Additionally, our theory contradicts the idea that the coherence time of the time crystal is limited by the decoherence or decay time of erbium ions. As long as the driving field and nonlinear interaction can disrupt time translation symmetry, the time crystalline order remains persistent. Our experiments also show that the self-organized periodic oscillations coexist with the intrinsic optical instability phase of erbium ions[27], meaning the time crystalline order persists even amidst irregularly oscillating instabilities as fast as ~50 MHz.

Driven dissipative systems are known to self-support periodic motion, as demonstrated in the work of Prigogine and others[47]. Our erbium system shares some essential parallels with conventional dissipative structures. Firstly, erbium ions in our system is subject to atomic decay and dechonerence, thus forming an open quantum system. The application of a coherent laser further drives the system into a far-from-equilibrium state. Secondly, nonlinear erbium interactions, that is, the excitation induced frequency shift, play a crucial role in amplifying small perturbations and driving the erbium ensemble into a new stable dissipative order. Thirdly, the multiple-level energy structure of erbium ions provides sufficient degrees of freedom for triggering the intrinsic competition between different optical transitions. Nonetheless, the periodic motion of erbium ions differs from processes like the Bérnard experiment and the Belousov-Zhabotinsky reaction, as the erbium time crystal is a many-body system governed by a coherent driving rather than an intensity input. As a result, the phase of the pump laser significantly affects the time crystal, as shown in Fig. 3 and Note 10 in SI. In addition, the conventional dissipative structures require an external source of noise, while in our quantum erbium ensemble, the noise is intrinsic and does not require a specific external reference.

In conclusion, we have demonstrated in both theory and experiment that time crystal is an inherent phase of nonequilibrium many-body systems. The realization of an inherent time crystal in an ensemble of erbium ions indicates that many-body interaction not only can self-induce a temporal crystalline order, but also can protect it against environmental decoherence. The self-organized periodic oscillations are persistent and have a coherence time beyond that of individual erbium ions. The results here pave the way to create states of matter in a strongly correlated system. In particular, it should be possible to use many-body interactions to create robust quantum superposition states with long coherence times. Such phases can also be extended to microwave-driven systems for potential applications such as quantum metrology and quantum memories.

**Methods:**

**Theoretical method:** Due to the complex nature of many-body systems, the parameters of $H_{\text{sys}}$ in Eq. (1), such as $\delta_{g,i}$ and $J_{i,j}$, vary from atom to atom. It is impractical to calculate the response of such a system in a full quantum way. To obtain a basic understanding of the phases of the non-equilibrium erbium ensemble, we first disregard the inhomogeneities by setting $\delta_{g,i} \equiv \delta_g$, $\Delta_{e,i} \equiv \Delta_e$, $\Omega_{ge}(\mathrm{r}_i) \equiv \Omega_{ge}$, that is to say, the detunings and the Rabi frequencies for all erbium ions are the same. We then apply the mean-field approximation to our system, which means that the system density matrix can be factorized $\rho_{\text{sys}} = \otimes_k \rho_k$, and the reduced density matrix of the $i$th atom is given by $\rho_i = \text{Tr}_{\neq i}(\rho_{\text{sys}})$. The homogeneous mean-field Hamiltonian of our system is written as

$$H_i = \delta_2 \sigma_{22} + \delta_3 \sigma_{33} + \delta_4 \sigma_{44} \\ + \Omega_{13}\sigma_{13} + \Omega_{14}\sigma_{14} + \Omega_{23}\sigma_{23} + \Omega_{24}\sigma_{24} \\ + \Delta_s(\rho_{44} + \rho_{33} - \rho_{22} - \rho_{11})\sigma_{33} + \Delta_s(\rho_{44} + \rho_{33} - \rho_{22} - \rho_{11})\sigma_{44}, \quad (2)$$

Using this Hamiltonian, we can compute the Lindblad equation $\dot{\rho} = L\rho$ and obtain the time response of a collection of erbium ions (more details in Supplementary Note 1).

**Optical measurements:** The optical measurements are performed on an Er:Y$_2$SiO$_5$ sample, which has 1000 ppm of yttrium ions replaced by erbium ions. The resonance frequency of the erbium ions is at 1536.4 nm (195177.17 THz), corresponding to spectroscopic site 1 (following the literature convention for spectroscopic site assignments). For notational convenience, all quoted laser frequencies $f_l$ in the paper are relative frequencies, that is, the absolute frequencies subtracted by 195117.17GHz.

One side of the sample was coated to achieve a reflectivity of 98.8 %, while the other end was coated with an anti-reflective layer of less than 0.8 % reflectivity. The sample is cooled to 4.0 K by a cryostat. The intensity instability of the of our laser (E15, NTK) is 2% in a 10s scale, and the laser linewidth is documented as less than 100 Hz.

**Data analysis using cross-correlation function:** A reference pulse function with an oscillating frequency of $\omega_r$ is defined as $r(t) = \cos(\omega_r t) \cdot \text{rect}(t)$, where

$$\text{rect}(t) = \begin{cases} 0, & \text{if } t < 0 \\ 1, & \text{if } 0 \leq t \leq T \\ 0, & \text{if } t > T \end{cases} \qquad (3)$$

is a rectangle function with an open duration of $T$. The cross-correlation function between our measured $I(t)$ and $r(t)$ is defined as $F(\tau) = \int_{-\infty}^{+\infty} I(t) r(t - \tau) dt$. It is already known that the phase noise $\phi(t)$ of $I(t)$ is a slowly-varying function of time (on the order of 1 ms), and the auto-correlation function of $I(t)$ indicates a periodicity at the frequency of 8.7 kHz. Note also that, as discussed in our previous work[30], $I(t)$ varies rapidly at frequencies on the order of 10 MHz. By choosing $T \sim 0.1$ ms and $\omega_r = 8.7$ kHz, which leads to the limit that $(\omega_0 - \omega_r)T \to 0$, we obtain that $F(\tau) \propto \cos[\omega_0 \tau + \phi(\tau)]$. This result means that we can reproduce the phase information $\phi(t)$ of $I(t)$ by choosing different $\tau$ when calculating $F(\tau)$. The time resolution is determined by $T$.

**Data Availability:** The experimental data supporting the findings in this paper are included in the main article and the Supplementary Materials. The codes can be obtained from the corresponding author upon request. The source data generated in this study are provided in the Source Data file.

**Acknowledgements:** The authors would like to thank Zhang-Qi Yin for helpful discussions. This work was supported by the National Natural Science Foundation of China (Nos. 62105033 and 12174026).

**Author Contributions Statement:** Y. H. C and X. Z. developed the idea, conducted the work and prepared the manuscript.

**Competing Interests Statement:** The authors declare no competing interests.

# Supplemental Information: Realization of an inherent time crystal in a dissipative many-body system

Yu-Hui Chen

Xiangdong Zhang

Key Laboratory of advanced optoelectronic quantum architecture and measurements of Ministry of Education,
Beijing Key Laboratory of Nanophotonics & Ultrafine Optoelectronic Systems,
School of Physics, Beijing Institute of Technology, 100081, Beijing, China

**CONTENTS:**

# Supplementary Note 1. System Hamiltonian and mean-field approximation

We consider a collection of four-level atoms whose energy structure consists of two pairs of Kramers doublets separated by an optical transition. The Hamiltonian of an ensemble of CW-pumped four-level atoms is given by

$$H_{\text{sys}} = \sum_i (\sum_{g,e} \delta_{g,i} |g\rangle\langle g|_i + \Delta_{e,i} |e\rangle\langle e|_i + \Omega_{ge}(\mathbf{r}_i)|g\rangle\langle e|_i + \Omega_{ge}(\mathbf{r}_i)|e\rangle\langle g|_i) \\ + \sum_{i,j} \frac{J_{ij}}{2|\mathbf{r}_{ij}|^3} [\mathbf{S}_i \cdot \mathbf{S}_j - 3(\mathbf{S}_i \cdot \hat{\mathbf{r}}_{ij})(\mathbf{S}_j \cdot \hat{\mathbf{r}}_{ij})], \quad (1)$$

where $|g\rangle_i$ and $|e\rangle_i$ indicate the optical ground and excited states of the $i$th atom, $g = 1,2$ and $e = 1,2$ means that there are a pair of Kramers doublets in the ground and excited states, $\delta_{g,i}$ is the ground-state detuning, $\Delta_{e,i}$ is the excited-state detuning, $\Omega_{ge}(\mathbf{r}_i)$ is the optical Rabi frequency of the $i$th atom, $J_{ij}$ is the strength of the magnetic dipole interactions, $\mathbf{r}_{i,j}$ is the vector connecting two spins $\mathbf{S}_i$ and $\mathbf{S}_j$, and $\hat{\mathbf{r}}_{ij}$ is the corresponding unit vector. The system Hamiltonian $H_{\text{sys}}$ is the sum of the Hamiltonian of individual ion $H_i$, which means that $H_{\text{sys}} = \sum_i H_i$ with

$$H_i = \sum_{g,e} \delta_{g,i} |g\rangle\langle g|_i + \Delta_{e,i} |e\rangle\langle e|_i + \Omega_{ge}(\mathbf{r}_i)|g\rangle\langle e|_i + \Omega_{ge}(\mathbf{r}_i)|e\rangle\langle g|_i \\ + \sum_{j \neq i} \frac{J_{ij}}{2|\mathbf{r}_{ij}|^3} [\mathbf{S}_i \cdot \mathbf{S}_j - 3(\mathbf{S}_i \cdot \hat{\mathbf{r}}_{ij})(\mathbf{S}_j \cdot \hat{\mathbf{r}}_{ij})]. \quad (2)$$

In principle, the parameters in $H_i$, such as $\delta_{g,i}$ and $J_{i,j}$, varies from atom to atom. It is impractical to calculate in a full quantum way the response of such a system. Here we consider that the atoms are homogeneous, i.e., $\delta_{g,i} \equiv \delta_g$, $\Delta_{e,i} \equiv \Delta_e$, $\Omega_{ge}(\mathbf{r}_i) \equiv \Omega_{ge}$. Then the Hamiltonian $H_i$ becomes

$$H_i = \sum_{g,e} (\delta_g|g\rangle\langle g| + \Delta_e|e\rangle\langle e| + \Omega_{ge}|g\rangle\langle e| + \Omega_{ge}|e\rangle\langle g|) + \sum_{j \neq i} \frac{V_{i,j}}{2}, \quad (3)$$

where the interaction term is rewritten as $V_{i,j}$. The dipole-dipole interactions $V_{i,j}$ in our theoretical model depends on the relative position $\mathbf{r}_{ij}$ of the ions and the relative orientation of the dipoles. This interaction presents two main features: a long-range character through the $1/r^3$ decay (instead of a short-range $1/r^6$ decay) and an anisotropic nature in space. The total interaction on the $i$th ion is the sum of the contribution of all the $j$th ions, i.e., $V_i = \sum_j V_{i,j}$. An intuitive way to see the long-range behaviour of such an interaction is to use an integral to replace the sum, i.e., $V_i = \int V(\mathbf{r}_{i,j}) d\mathbf{r}_j$. As the number of interacting ions with similar interaction strength $V(\mathbf{r}_{i,j}) = V(r) \propto 1/r^3$ grows with $r^2$. Then the total interaction strength on the $i$th ion $V_i$ can be estimated by an integral of the form of $\int r^2 \cdot 1/r^3 dr$. This equation suggests that the integral grows with increasing $r$ and even become divergent when the crystal size $r \to \infty$. Therefore, the dipole-dipole interaction in our sample is long range. To precisely calculate $V_i$, a more dedicate model involving the microscopic information of the doped erbium ions is needed. Here, we apply the mean-field approximation to study the influence of the term $V_i$. The mean-field theory means that the density matrix of the whole system can be factorized $\rho_{\text{sys}} = \otimes_k \rho_k$, and the reduced density matrix of the $i$th atom is $\rho_i = \text{Tr}_{\neq i}(\rho_{\text{sys}})$.

The equation of motion for the system is $\dot{\rho}_{\text{sys}} = -i[\rho_{\text{sys}}, H_{\text{sys}}]$. Then the dynamic equation of the reduced $\rho_i$ is

$$\begin{aligned}\dot{\rho}_i &= \frac{d}{dt}[\text{Tr}_{\neq i}(\rho_{\text{sys}})] \\ &= \text{Tr}_{\neq i}(\dot{\rho}_{\text{sys}}) \\ &= -i\,\text{Tr}_{\neq i}(\rho_{\text{sys}} H_{\text{sys}} - H_{\text{sys}} \rho_{\text{sys}}) \\ &= -i\,\text{Tr}_{\neq i}\left(\rho_{\text{sys}} \sum_m H_m - \sum_m H_m\, \rho_{\text{sys}}\right).\end{aligned} \quad (4)$$

Note that $H_m$ is given by Eq. (3)

$$\begin{aligned}H_m &= \sum_{g,e}(\delta_g |g\rangle\langle g| + \Delta_e |e\rangle\langle e| + \Omega_{ge}|g\rangle\langle e| + \Omega_{ge}|e\rangle\langle g|) + \sum_{n \neq m}\frac{V_{m,n}}{2} \\ &\equiv h_m + \sum_{n \neq m}\frac{V_{m,n}}{2},\end{aligned} \quad (5)$$

where $h_m$ contains the detuning terms and the optical driving terms, and $V_{m,n}$ stands for the many-body interactions. For the first term $h_m$ in Eq. (5), we then obtain

$$\mathrm{Tr}_{\neq i}\left(\rho_{\mathrm{sys}}\sum_m h_m - \sum_m h_m\,\rho_{\mathrm{sys}}\right) = \mathrm{Tr}_{\neq i}\left(\otimes\rho_k \sum_m h_m - \sum_m h_m \otimes\rho_k\right)$$

$$= \mathrm{Tr}_{\neq i}\{\sum_m [(\rho_m h_m - h_m \rho_m)\bigotimes_{k\neq m}\rho_k]\}$$

$$= \mathrm{Tr}_{\neq i}[(\rho_i h_i - h_i \rho_i)\bigotimes_{k\neq i}\rho_k] + \mathrm{Tr}_{\neq i}[\sum_{m\neq i}(\rho_m h_m - h_m \rho_m)\bigotimes_{k\neq m}\rho_k] \quad (6)$$

$$= (\rho_i h_i - h_i \rho_i)\mathrm{Tr}_{\neq i}\left(\bigotimes_{l\neq i}\rho_l\right) - 0$$

$$= (\rho_i h_i - h_i \rho_i),$$

where we have used $\mathrm{Tr}_{\neq i}(\otimes_{k\neq i}\rho_k) = 1$ and $\mathrm{Tr}(\rho_m h_m - h_m \rho_m) = 0$. This equation means a well-known result that $\dot{\rho}_i = -i[\rho_i, h_i]$ when there is no cross talk $V_{m,n}$ between different atoms.

For the second term $V_{m,n}$ in Eq. (5), following a similar procedure, we obtain

$$\begin{aligned}
\mathrm{Tr}_{\neq i}\left(\rho_{\mathrm{sys}}\sum_{m,n} V_{m,n} - \sum_{m,n} V_{m,n}\,\rho_{\mathrm{sys}}\right) &= \mathrm{Tr}_{\neq i}\left(\otimes\rho_k \sum_{m,n} V_{m,n} - \sum_{m,n} V_{m,n} \otimes\rho_k\right) \\
&= \mathrm{Tr}_{\neq i}\{\sum_{m,n}[(\rho_m \otimes \rho_n \cdot V_{m,n} - V_{m,n}\cdot \rho_m \otimes \rho_n)\bigotimes_{k\neq m,n}\rho_k]\} \\
&= \mathrm{Tr}_{\neq i}\left[\left(\sum_n \rho_i \otimes \rho_n \cdot V_{i,n} - \sum_n V_{i,n}\cdot \rho_i \otimes \rho_n\right)\bigotimes_{k\neq i,n}\rho_k\right] + \\
&\quad \mathrm{Tr}_{\neq i}\left[\left(\sum_m \rho_m \otimes \rho_i \cdot V_{m,i} - \sum_m V_{m,i}\cdot \rho_m \otimes \rho_i\right)\bigotimes_{k\neq i,m}\rho_k\right] + \\
&\quad \mathrm{Tr}_{\neq i}\left[\left(\sum_{m\neq i,n\neq i} \rho_m \otimes \rho_n \cdot V_{m,n} - \sum_{m\neq i,n\neq i} V_{m,n}\cdot \rho_m \otimes \rho_n\right)\bigotimes_{k\neq m,n,i}\rho_k\right]\cdot \rho_i.
\end{aligned} \quad (7)$$

For the first term in the above equation, we have

$$\begin{aligned}
\mathrm{Tr}_{\neq i}\left[\left(\sum_n \rho_i \otimes \rho_n \cdot V_{i,n} - \sum_n V_{i,n}\cdot \rho_i \otimes \rho_n\right)\bigotimes_{k\neq i,n}\rho_k\right] &= \mathrm{Tr}_n\left[\left(\sum_n \rho_i \otimes \rho_n \cdot V_{i,n} - \sum_n V_{i,n}\cdot \rho_i \otimes \rho_n\right)\right]\cdot \mathrm{Tr}_{\neq i,n}\left(\bigotimes_{k\neq i,n}\rho_k\right) \\
&= \mathrm{Tr}_n\left[\left(\sum_n \rho_i \otimes \rho_n \cdot V_{i,n} - \sum_n V_{i,n}\cdot \rho_i \otimes \rho_n\right)\right] \quad (8)\\
&= \rho_i \otimes \mathrm{Tr}_n\left(\sum_n \rho_n \cdot V_{i,n}\right) - \mathrm{Tr}_n\left(\sum_n V_{i,n}\cdot \rho_n\right)\otimes \rho_i.
\end{aligned}$$

Similarly, the second term in Eq. (7) becomes

$$\mathrm{Tr}_{\neq i}\left[\left(\sum_m \rho_m \otimes \rho_i \cdot V_{m,i} - \sum_m V_{m,i}\cdot \rho_m \otimes \rho_i\right)\bigotimes_{k\neq i,m}\rho_k\right] = \rho_i \otimes \mathrm{Tr}_m\left(\sum_m \rho_m \cdot V_{m,i}\right) - \mathrm{Tr}_m\left(\sum_m V_{m,i}\cdot \rho_m\right)\otimes \rho_i. \quad (9)$$

Using again that $\mathrm{Tr}(AB - BA) = 0$, we obtain for the third term in Eq. (7) that

$$\text{Tr}_{\neq i}\left[\left(\sum_{m\neq i, n\neq i}\rho_m \otimes \rho_n \cdot V_{m,n} - \sum_{m\neq i, n\neq i} V_{m,n} \cdot \rho_m \otimes \rho_n\right)\bigotimes_{k\neq m,n}\rho_k\right] = 0. \tag{10}$$

Substituting Eq. (8), (9) and (10) into Eq. (7), we have

$$\frac{1}{2}\text{Tr}_{\neq i}\left(\rho_{\text{sys}}\sum_{m,n}V_{m,n} - \sum_{m,n}V_{m,n}\,\rho_{\text{sys}}\right) = \rho_i \otimes \text{Tr}_n\left(\sum_n \rho_n \cdot V_{i,n}\right) - \text{Tr}_n\left(\sum_n V_{i,n}\cdot \rho_n\right)\otimes \rho_i. \tag{11}$$

By expanding $V_{i,n}$ in the basis of $|g\rangle$ and $|e\rangle$, we can rewrite it as

$$V_{i,n} = \sum_{g_i, g_n, e_i, e_n} \langle g_i g_n|V_{i,n}|g_i g_n\rangle \cdot |g_i g_n\rangle\langle g_i g_n| + \langle g_i e_n|V_{i,n}|g_i e_n\rangle \cdot |g_i e_n\rangle\langle g_i e_n| \\
+ \langle e_i g_n|V_{i,n}|e_i g_n\rangle \cdot |e_i g_n\rangle\langle e_i g_n| + \langle e_i e_n|V_{i,n}|e_i e_n\rangle \cdot |e_i e_n\rangle\langle e_i e_n|. \tag{12}$$

In the above expansion, those terms as $|g_i g_n\rangle\langle g_i e_n|$ are dropped under the rotating wave approximation, and the energy transfer terms as $|g_i e_n\rangle\langle e_i g_n|$ are assumed to be negligible. Substituting Eq. (11) into the tracing $\text{Tr}_n(\sum_n \rho_n \cdot V_{i,n})$ in Eq. (12) gives

$$\text{Tr}_n\left(\sum_n \rho_n \cdot V_{i,n}\right) = \sum_{g_i, g_n, e_i, e_n} \langle g_i g_n|V_{i,n}|g_i g_n\rangle \cdot \rho_{gg,n}\cdot|g\rangle\langle g|_i + \langle g_i e_n|V_{i,n}|g_i e_n\rangle \cdot \rho_{ee,n}\cdot|g\rangle\langle g|_i \\
+ \langle e_i g_n|V_{i,n}|e_i g_n\rangle \cdot \rho_{gg,n}\cdot|e\rangle\langle e|_i + \langle e_i e_n|V_{i,n}|e_i e_n\rangle \cdot \rho_{ee,n}\cdot|e\rangle\langle e|_i \\
\stackrel{\text{def}}{=} \sum_{g_i, g_n, e_i, e_n} V_a \cdot \rho_{gg,n}|g\rangle\langle g|_i + V_b \cdot \rho_{ee,n}|g\rangle\langle g|_i + V_c \cdot \rho_{gg,n}|e\rangle\langle e|_i + V_d \cdot \rho_{ee,n}|e\rangle\langle e|_i, \tag{13}$$

where the coefficients $V_a = \langle g_i g_n|V_{i,n}|g_i g_n\rangle$ and so do $V_b$, $V_c$ and $V_d$. It is clear that the effect of $V_{m,n}$ is to shift the energy of the states $|g\rangle$ and $|e\rangle$. Generally, the energy shifts for the ground and excited states are not the same. In optical experiments where a laser field drives the transitions between the states of $|g\rangle$ and $|e\rangle$, what matters is the frequency differences between them. Therefore, the induced frequency difference $\chi$ of the optical ground and excited states of the $i$th ion becomes

$$\chi_n = \sum_{g_n, e_n} V_c\,\rho_{gg,n} + V_d\,\rho_{ee,n} - V_a\,\rho_{gg,n} - V_b\,\rho_{ee,n} \\
= \sum_{g_n, e_n} \frac{V_c + V_d - V_a - V_b}{2}(\rho_{gg,n} + \rho_{ee,n}) + \frac{V_a + V_d - V_b - V_c}{2}(\rho_{ee,n} - \rho_{gg,n}). \tag{14}$$

Noting that $\sum_{g,e} \rho_{gg,n} + \rho_{ee,n} = 1$, we can neglect the first term in the above equation as it simply implies a constant frequency shift. Then we obtain

$$\chi_n = \sum_{g_n,e_n} \eta_n (\rho_{ee,n} - \rho_{gg,n}), \tag{15}$$

where $\eta_n = (V_a + V_d - V_b - V_c)/2$. It is now clear that the effect of ion-ion interactions $V_{m,n}$ is to introduce an excitation-dependant frequency shift to the $i$th ion. With Eq. (3) and ([eq.vij_reduced]), we can write the Hamiltonian of the $i$th erbium ion as

$$H_i = h_i + \sum_n V_{i,n}$$

$$= \sum_{g_i,e_i} \delta_g |g\rangle\langle g| + \Delta_e |e\rangle\langle e| + \Omega_{ge} |g\rangle\langle e| + \Omega_{ge} |e\rangle\langle g|_i + \sum_n \sum_{g_n,e_n,e_i} \eta_n (\rho_{ee,n}$$

$$- \rho_{gg,n}) |e\rangle\langle e|_i. \tag{16}$$

Hamiltonians of a similar form as Eq. (16) are widely used to analyze phase transitions in electron spins, Rydberg atoms, and rare-earth ions. For a uniform version of Eq. , that is, the density matrix is the same for all ions ($\rho_{ee,i} = \rho_{ee,n} \equiv \rho_{ee}$ and $\rho_{gg,i} = \rho_{gg,n} \equiv \rho_{gg}$), we can obtain

$$H_i = \sum_{g,e} (\delta_g |g\rangle\langle g| + \Delta_e |e\rangle\langle e| + \Omega_{ge} |g\rangle\langle e| + \Omega_{ge} |e\rangle\langle g| + \Delta_s \sum_{g',e'} (\rho_{e'e'} - \rho_{g'g'}) |e\rangle\langle e| \ ), \tag{17}$$

where $\Delta_s = \sum_n \eta_n$ is the excitation-induced frequency shift. This kind of frequency shift is often considered to be a decoherence source and recently be employed to demonstrate quantum gate operations and conditional quantum phase shift. Moreover, such a nonlinear effect is also the origin of intrinsic optical intrinsic optical instabilities. Rewriting Eq. (17) explicitly for a four-level system, we have

$$\begin{aligned} H_i = \ & \delta_2 \sigma_{22} + \delta_3 \sigma_{33} + \delta_4 \sigma_{44} \\ + \ & \Omega_{13} \sigma_{13} + \Omega_{14} \sigma_{14} + \Omega_{23} \sigma_{23} + \Omega_{24} \sigma_{24} \\ + \ & \Delta_s (\rho_{44} + \rho_{33} - \rho_{22} - \rho_{11}) \sigma_{33} + \Delta_s (\rho_{44} + \rho_{33} - \rho_{22} - \rho_{11}) \sigma_{44}. \end{aligned} \tag{18}$$

With this Hamiltonian, we can now compute the Lindblad equation $\dot{\rho} = L\rho$ by using the Runge-Kutta method (Notre 2 or more detailed in our previous work). In our experiment, the measured light intensity $I(t)$ is related to $\rho(t)$ through the macroscopic polarization $P(z,t)$. We consider a laser field $\mathcal{E}(z,t) = E(z,t)\exp(-i\omega t + ikz + \phi)$ passing through our sample along the $z$ direction from $z = 0$ to $z = L$, where $L$ is the length of our sample. The intensity $I(t)$ is proportional to $|E(z = L, t)|^2$. Applying the slowly-varying-amplitude approximation to $E(z,t)$, we have

$$\frac{\partial}{\partial z} E(z,t) = \frac{k}{2\epsilon} \cdot \text{Im} P(z,t), \tag{19}$$

where $k$ is the wave vector, $\epsilon$ is the permittivity constant, and $\text{Im} P(z,t)$ is the imaginary part of the polarization $P(z,t)$. Given the density of erbium ions $n_{\text{er}}$, the polarization $P(z,t)$ is related to the above-calculated $\rho(t)$ by

$$P(z,t) = n_{\text{er}}[d_{12} \cdot \rho_{12}(z,t) + d_{13} \cdot \rho_{13}(z,t) + d_{23} \cdot \rho_{23}(z,t) + d_{24} \cdot \rho_{24}(z,t)]. \tag{21}$$

We then obtain

$$E(L,t) = \int_0^L dz \cdot \frac{n_{\text{er}} k}{2\epsilon} \cdot Im[d_{12} \cdot \rho_{12}(z,t) + d_{13} \cdot \rho_{13}(z,t) + d_{23} \cdot \rho_{23}(z,t) + d_{24} \cdot \rho_{24}(z,t)] \tag{22}$$

Since we are only interested in the temporal response of our system, Eq. (21) clearly shows that an oscillating $\rho(t)$ can result in a oscillating $E(L,t)$ and $I(t)$.

Note that the Hamiltonian Eq. (18) is based on a homogeneous model. However, erbium ions doped in crystals are subject to inhomogeneous broadening. This means that the frequencies of their optical transitions and spin transitions can vary from ion to ion within the crystal. The inhomogeneous broadening of optical transition is approximately 1GHz and that of spin transitions is on the order of 10MHz. In our system, the absolute optical transition frequency is not important as it can be compensated by tuning the laser frequency. What really matters is the detuning terms such as $\delta_2$, $\delta_3$ and $\delta_4$. Applying a laser of mw power to drive our erbium ensemble results in approximately 10% of the ion in the inhomogeneous line are excited. This implies that the $\delta_2$, $\delta_3$ and $\delta_4$ corresponding to the excited ions vary within a range of ~1MHz. We thus assuming that the atomic detunings are the same in our theoretical model. Note that in our experiments, some ions with different frequency detunings can also be excited. However, they are likely not in a dynamically stable phase and can just impose a cw background in our time-crystal detection.

## Supplementary Note 2. Many body interactions and time crystalline order

**I. Many-body interations**

The Hamiltonian Eq. 18 can be represented as a matrix and consists of three parts:

$$H = H_0 + H_r + H_m \tag{23}$$

where

$$H_0 = \begin{bmatrix} 0 & 0 & 0 & 0 \\ 0 & \delta_2 & 0 & 0 \\ 0 & 0 & \delta_3 & 0 \\ 0 & 0 & 0 & \delta_4 \end{bmatrix}, \tag{24}$$

represents the Hamiltonian of free ions,

$$H_r = \begin{bmatrix} 0 & 0 & \Omega t_1 & \Omega t_2 \\ 0 & 0 & \Omega t_2 & \Omega t_1 \\ \Omega t_1 & \Omega t_2 & 0 & 0 \\ \Omega t_2 & \Omega t_1 & 0 & 0 \end{bmatrix}, \tag{24}$$

is the Rabi oscillation term governed by the optical driving, and

$$H_m = \begin{bmatrix} 0 & 0 & 0 & 0 \\ 0 & 0 & 0 & 0 \\ 0 & 0 & \Delta_s(\rho_{44} + \rho_{33} - \rho_{22} - \rho_{11}) & 0 \\ 0 & 0 & 0 & \Delta_s(\rho_{44} + \rho_{33} - \rho_{22} - \rho_{11}) \end{bmatrix}, \tag{25}$$

represents the many-body interactions.

The expression of $H_m$ indicates that when the input laser is weak (or absent), most erbium ions remain in their ground states. In this case, $H_m$ acts as a constant frequency shift on the states $|3\rangle$ and $|4\rangle$. However, as more ions are excited and the values of $\rho_{33}$ and $\rho_{44}$ become significant, the frequencies of the

targeted ions are altered. This kind of many-body interactions serves as the underlying mechanism for various quantum phenomena, including quantum gate operations, intrinsic optical bistability, and intrinsic optical instability.

With the above Hamiltonian and the density matrix of an erbium ion

$$\rho = \begin{bmatrix} \rho_{11} & \rho_{12} & \rho_{13} & \rho_{14} \\ \rho_{21} & \rho_{22} & \rho_{23} & \rho_{24} \\ \rho_{31} & \rho_{32} & \rho_{33} & \rho_{34} \\ \rho_{41} & \rho_{42} & \rho_{43} & \rho_{44} \end{bmatrix}, \tag{26}$$

one can compute the time evolution of $\rho(t)$ through the Lindblad equation

$$\frac{d}{dt}\rho = -i[H,\rho] + Loss, \tag{27}$$

The $Loss$ term accounts for all the decay and dephasing processes of the erbium ions. It encompasses various rates, including the spin relaxation rate $\gamma_{12}$ between |2⟩ and |1⟩, the spin relaxation rate $\gamma_{34}$ between |4⟩ and |3⟩, the optical spontaneous emission rate $\gamma_{31}$ ($\gamma_{32}$) from |3⟩ to |1⟩ (|2⟩), the optical spontaneous emission rate $\gamma_{41}$ ($\gamma_{42}$) from |4⟩ to |1⟩ (|2⟩), the spin dephasing rates $\gamma_{22}$, $\gamma_{33}$, $\gamma_{44}$ of |2⟩, |3⟩ and |4⟩, respectively. These loss terms in the Lindblad form can be expressed as

$$L(\sigma_{jk}) = \frac{\gamma_{jk}}{2}(n_{jk}+1) \cdot (2\sigma_{jk}\rho\sigma_{kj} - \sigma_{kj}\sigma_{jk}\rho - \rho\sigma_{jk}\sigma_{kj}) + \frac{\gamma_{jk}}{2}n_{jk} \cdot (2\sigma_{kj}\rho\sigma_{jk} - \sigma_{jk}\sigma_{kj}\rho - \rho\sigma_{kj}\sigma_{jk}), \tag{28}$$

where $n_{jk}$ is the average photon number of the thermal bath at the frequency between level $j$ and $k$.

The parameters used in obtaining the results in Fig. 1 of the manucript are listed below. The detunings are $\delta_2 = 0.05$ MHz, $\delta_3 = -0.35$ MHz, and $\delta_4 = 0.4$ MHz. The induced-frequency shift is $\Delta_s = 12$ MHz. The dephasing rates are $\gamma_{22} = \gamma_{33} = \gamma_{44} = 1$kHz. The optical coupling strengths of |1⟩ to |3⟩ and that of |1⟩ to |4⟩ are $\Omega t_1$ and $\Omega t_2$, respectively, where $\Omega = 0.26$ MHz, $t_1 = 1.87$ and $t_2 = 1.2$. Similarly, the optical coupling strengths of |2⟩ to |3⟩ and that of |2⟩ to |4⟩ are $\Omega t_2$ and $\Omega t_1$, respectively. The optical excited states have a lifetime of $1/\gamma = 11$ms. The spin relaxation time for |2⟩ to |1⟩ and |4⟩ to |3⟩ are chosen to be 2 seconds. The longest time scale of the erbium dynamics in our simulations is 0.17

s, equivalent to 15 times the optical excited state lifetime of erbium ions. The oscillation depicted in Fig. 1 remains clear and unattenuated throughout the entire simulation period.

The $H_m$ in Eq. (18) represents the many-body interactions, which play a crucial role in generating the time crystalline order in our system. Without these interactions, self-sustained oscillations would be absent. To illustrate this effect, we computed the time evolution of $\rho(t)$ for various values of $\Delta_s$. Figure S1 shows the calculated $\rho_{33}(t)$ for different $\Delta_s$. When there are no ion-ion interactions ($\Delta_s = 0$ MHz), Rabi oscillations on the order of MHz can be observed shortly after the driving field is switched on (inset of Fig. S1(a)). However, the system quickly stabilizes at approximately 5 ms, indicating the absence of time crystalline order. If we include ion-ion interactions in the model with a small magnitude ($\Delta_s = 4$ MHz), the $\rho_{33}(t)$ remains stationary in the long time limit. When $\Delta_s$ is increased to 8 MHz, the $\rho_{33}(t)$ in the long time limit becomes dynamically unstable, indicating the spontaneous breaking of the continuous time translation symmetry. Further increasing $\Delta_s$ to 10 MHz leads to an even more pronounced instability of $\rho_{33}(t)$, further confirming the effect of the many-body interactions. The many-body interactions between erbium ions act as nonlinear intrinsic feedback, amplifying small perturbations and driving the erbium ensemble into a new stable dissipative order. Without these many-body interactions, no temporal instability or periodicity can be generated in our system. These results are consistent with the results reported in our previous publications.

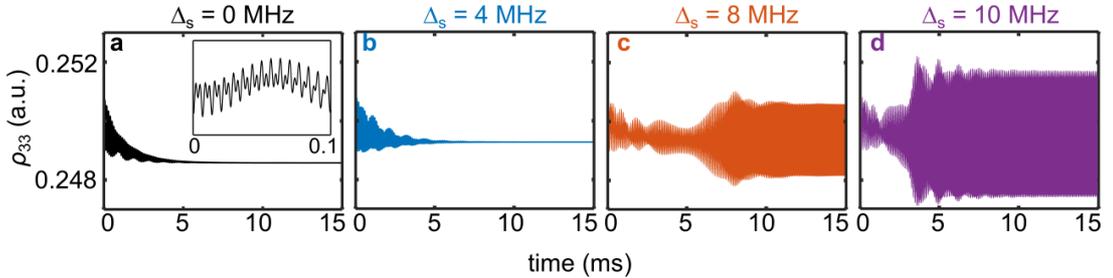

FIG. S1. **Population $\rho_{33}(t)$ for different many-body interactions $\Delta_s$.** (**a**)-(**d**), calculated population $\rho_{33}(t)$ for different $\Delta_s$ as noted. When the time translation invariance of $\rho_{33}(t)$ is broken for $\Delta_s = 8$ MHz and 10 MHz, Rabi oscillations can also be observed throughout the entire time range.

## II. Time crystalline order

When the time translation invariance of $\rho(t)$ is broken in the long time limit due to the inclusion of many-body interactions, Rabi oscillations induced by the optical driving can still be observed throughout the entire time range. This suggests that the system continues to undergo coherent population oscillations between the optical ground and excited states. More importantly, under appropriate parameters, the unstable $\rho(t)$ exhibits self-sustained oscillations at frequencies different from the Rabi oscillation frequencies. These regular oscillations indicate the formation of temporal order. For example, in Fig. S2(a), the Fourier spectrum of the calculated $\rho(t)$ (Fig. 1C in the main text) shows a peak at 46.4 kHz, confirming the presence of temporal order.

Within the large parameter space of the four-level systems, it is found that the characteristic frequency of the inherent time crystal is sensitive to the optical transition overlaps. Specifically, we set the optical coupling strengths of $|1\rangle$ to $|3\rangle$ and that of $|1\rangle$ to $|4\rangle$ to be $\Omega t_1$ and $\Omega t_2$, respectively, as shown in Fig.S2b. Here $\Omega$ depends on the pump laser, and $t_1$ and $t_2$ are coupling coefficients. Similarly, the optical coupling strengths of $|2\rangle$ to $|3\rangle$ and that of $|2\rangle$ to $|4\rangle$ are $\Omega t_2$ and $\Omega t_1$, respectively. After obtaining the time response of $\rho(t)$, we then carry out a Fourier transform to $\rho(t)$ in the long time scale. The time-crystal frequency as a function of the ratio of $t_1/t_2$ is plotted in Fig. S2c. For a range of $t_1/t_2$ varying from 1.08 to 1.20, the crystal frequency changes from 8.3 to 20.7kHz. The result suggests a strong dependence of the crystal frequency on the ratio of $t_1/t_2$.

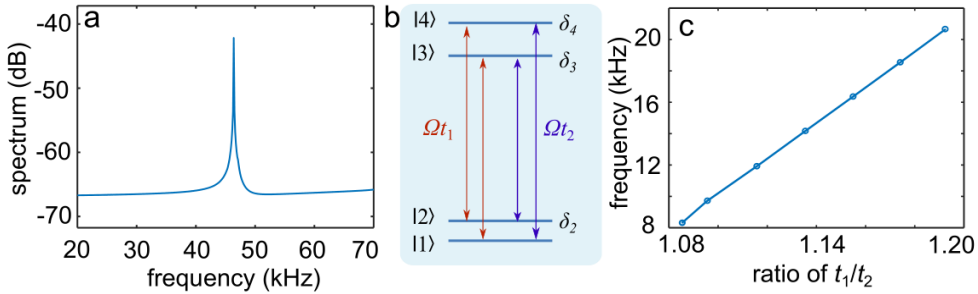

FIG. S2. **Time-crystal frequency as a function of $t_1/t_2$.** (**a**) The Fourier spectrum of $\rho_{11}(t)$ in Fig. 1C in the main text. (**b**) The four-level energy structure of an erbium ion. The transition strength between $|1\rangle$ and $|3\rangle$ and that between $|2\rangle$ and $|4\rangle$ are $\Omega t_1$. The transition strength between $|1\rangle$ and $|4\rangle$ and between $|2\rangle$ and $|3\rangle$ are $\Omega t_2$. (**c**) The time-crystal frequency as a function of the ratio $t_1/t_2$.

The physics of such dependence can be understood from Fig. 1D and 1E in the main text. As discussed, it is essential to have enough complexity in the energy structure to enable competition between different optical transitions. Such competition is necessary to bring an ensemble of atoms into a dynamically unstable phase. For pure two-level systems, the lack of complexity prevents such competition between different optical transitions and thus rules out a temporal-order phase. For four-level systems, the ratio $t_1/t_2$ indicates the overlaps between the different transitions. Therefore, it is an important index of the competing processes and can significantly affect the time crystal frequency.

In contrast, our calculated results suggest the lifetime or the decoherence time of individual erbium ions has no significant impact on the time scale of the crystalline oscillation. The dissipation terms of erbium ions are more important in determining whether a time crystal phase can be form. While a too-low damping rate might not be enough to efficiently expel the heating due to the dynamic instability, a high rate can cause the rapid dephasing of the oscillations.

It is worth noting that the emergence of time crystalline order in our four-level system is not simply a result of combining different Rabi oscillations. As shown in Figure S1, without the presence of many-body interactions, these Rabi oscillations would only last for a short period of time and eventually fade out, indicating that the time crystalline behaviour cannot be solely explained by Rabi oscillations. Furthermore, if the time crystal were solely the effect of Rabi oscillations, we would expect the crystal frequency to be affected by both the Rabi frequency and the ratio of transitions, which contradicts our calculations and experimental results.

## Supplementary Note 3. Feedback loop of the four-level system

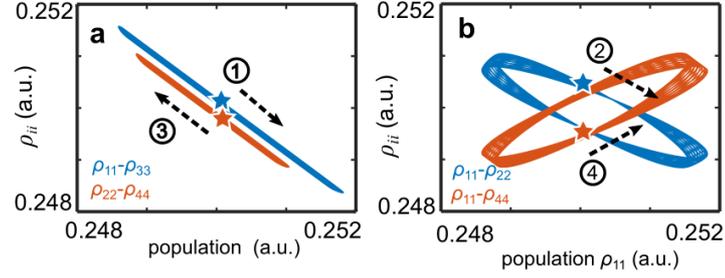

FIG. S3. **Feedback loop.** If there is a small perturbation to $\rho_{11}$, the system behaves as positive feedback following the steps as noted by the circled numbers.

We use the calculated results in Fig. 1 in the main text to demonstrate the feedback loop of the four-level system. Starting for the steady solution corresponding to $\dot{\rho}(t) = 0$, we assume that there is a small positive perturbation to $\rho_{11}$, as marked by in Fig. S3a. According to Fig. S3b, such an increase in $\rho_{11}$ will lead to a decrease of $\rho_{22}$, as marked by . Then the decrease of $\rho_{22}$ will result in a growth of $\rho_{44}$, as marked by in Fig. S3a. Finally, the increased $\rho_{44}$ further enhances the increase of $\rho_{11}$ as marked by in Fig. S3b. These processes form intrinsic positive feedback inside the four-level system. The positive feedback, together with the optical transitions that compete with each other, leads the system to a dynamical instability and a self-organized periodic temporal pattern.

The limit cycles as shown in S3 is a typical sign of the breaking of time translation symmetry. In the atom-cavity time crystal , if the coupling between the cavity field and the atomic polarization is linear, the system relaxes to a stationary state at the long-time limit. The limit-cycle behaviours therein are the result of the nonlinear interaction between the cavity field and the atomic polarization. In our system, the nonlinear interactions between different electronic transitions offer the possibility of breaking time translation symmetry. The two systems share similarity in math representations, as both the cavity field and the electronic transition can be represented by an oscillator in math.

# Supplementary Note 4. Dipole-dipole interactions of erbium ions

Our Er:Y$_2$SiO$_5$ crystal has a concentration of 1000 ppm, corresponding to an average distance of 4nm between nearby erbium ions. The magnetic dipole-dipole interaction of erbium ions separated by 4nm is $\mathcal{O}(10\text{MHz})$. Note that erbium ions also possess electric dipole moments. It is estimated that the electric dipole induced Stark-shift of our system is also $\mathcal{O}(10\text{MHz})$. Thus the contributions to the excitation-induced-frequency-shift $\Delta_s$ in Eq. (S.18) from the magnetic and the electric interactions may have a comparable magnitude. However, it is not important in our case whether the $\Delta_s$ are magnetic or electric. These two types of interactions have similar mathematical forms and the same nonlinear effect on our system. Without losing generality, we here considered the many-body erbium interactions being magnetic.

Our theory predicts that a collection of four-level atoms with dipole-dipole interactions can give rise to an inherent time crystalline phase. When choosing a proper material to realize inherent time crystal, it is important to make sure the many-body interactions play a role. Therefore the ratio between the dipole-dipole interactions, optical Rabi frequency, and the dephasing rate should be high. Otherwise, the many-body interactions can not be distinguished from the background consisting of various decoherence processes. Although the erbium-erbium interactions are not as strong as that of Rydberg atoms, erbium-doped crystals feature extremely long optical and spin coherence times. Therefore the dipole-dipole interactions between nearby erbium ions are experimentally distinguishable, providing an experimental platform of investigating the formation for inherent time crystals.

# Supplementary Note 5. Long time behaviors

Figure 2 of the main text shows the spontaneously breaking of continuous time translation symmetry and the forming of temporal order in a time scale of 20ms. In our experiment, the oscillating $I(t)$ lasts for ever as long as the pump laser is on. This is in contrast to the continuous time crystal recently demonstrated, which has a lifetime limited by the atom loss. The persistence of $I(t)$ in our experiment is explicitly shown in Fig. S4, where the $I(t)$ after the laser ~100ms after the switching of the laser is shown. As long as the pump laser is on its cw mode, such oscillations can be observed. This property is crucial as it enables us to monitor the spectra of $I(t)$ in the long-time limit (or cw mode), as shown in Fig. 3 in the main text.

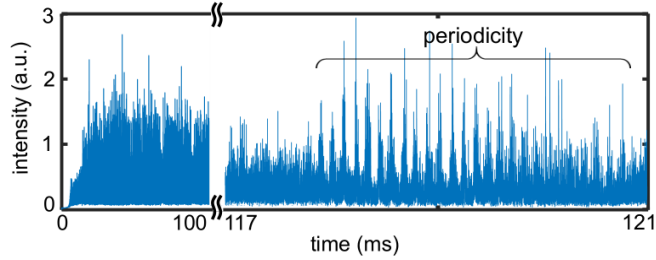

FIG. S4. **Long time behaviors of the system.** An example of $I(t)$ measured from $t = 0$ ms to $t = 121$ ms. Note that the $I(t)$ of the first 100 ms and the $I(t)$ from 117 ms to 120 ms have different scales on their x-axes.

## Supplementary Note 6. Phase diagram

Figure S5 shows the phase diagram of our system. Normally the output $I(t)$ of our measurement is dynamically stable for different combination of pump power $P_{\text{in}}$ and laser frequency $f_l$. Starting from a cw state, the output of our system under increasing $P_{\text{in}}$ first becomes dynamically unstable, then temporally periodic and finally dynamically irregular again.

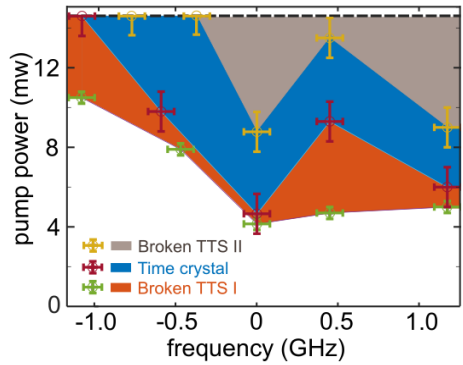

FIG. S5. **Phase diagram of time crystal as a function of laser frequency and pump power.** TTS, time translation symmetry. White area, a phase with stable cw ouput; blue area, the phase of Broken TTS I, where the output becomes dynamically unstable, but without a distinguishable periodic signal; orange area, the phase of time crystal, where a 8.7 kHz periodic signal can be distinguished from the background; gray area, the phase of Broken TTS I, where the output remains dynamically unstable but the 8.7 kHz cycles no longer can be identified due to the increased oscillating background. The cross points are experimentally measured points.

With increasing pump, the $I(t)$ first becomes dynamically unstable, indicating the breaking of continuous time translation symmetry. The spectral range of $I(t)$ is broad with a cut-off frequency $\sim 50$ MHz, which means that $I(t)$ oscillates at the time scale of 10ns. However, no periodic signal can be identified from the correlation $\langle I(t)I(t+\tau)\rangle$ or the spectra of $I(t)$. In other words, even though the continuous time translation symmetry of our system is broken, there is lack of a temporal order. This regime is named as the phase of broke time translation symmetry I, as shown by the orange area in Fig. S5.

Depending on the $f_l$, if the $P_{\text{in}}$ is further increased, the many-body system can reach a phase with temporal order. For example, for $f_l = 0.00$ GHz, if $P_{\text{in}}$ is increased to 5.3mW, a periodic oscillation of 8.7KHz can be identified in the already-unstable $I(t)$. For $f_l = 0.50$ GHz and $f_l = 1.12$ GHz, a pump of $P_{\text{in}} = 9.3$mW and 6.0mW are needed, respectively (dark-red cross points in Fig. S5). Note that under this circumstance, the $I(t)$ is still dynamically unstable and the MHz oscillating components remain there. This regime features the broken time translation symmetry together with a self-generated periodicity, which is named as the phase of time crystal, as shown by the blue area in Fig. S5.

Further increasing $P_{\text{in}}$ can lead to the fading of the time crystalline signal. The reason is twofold. On one hand, the 8.7kHz peak becomes broadened for increasing $P_{\text{in}}$ (as shown in Fig. 3a in the manuscript); on the other hand, the oscillation components of $I(t)$ at other frequencies grows with $P_{\text{in}}$ (as shown in Note 7, SI). Both the two factors make the periodic signal less distinguishable. Note also that although the time crystalline order fades, the $I(t)$ is still dynamically unstable. This regime is named as the phase of broke time translation symmetry II, as shown by the gray area in Fig. S5.

Note that there are some experimental conditions that limit the determination of the phase transition points. Because the moment when the cw output becomes dynamically unstable, and the moment when the 8.7kHz can be distinguishable from the background, are lack of a strict standard and rely on experimental experience. Therefore, there are some errors when plotting the phase transition points.

# Supplementary Note 7. Phase transitions at different laser frequencies

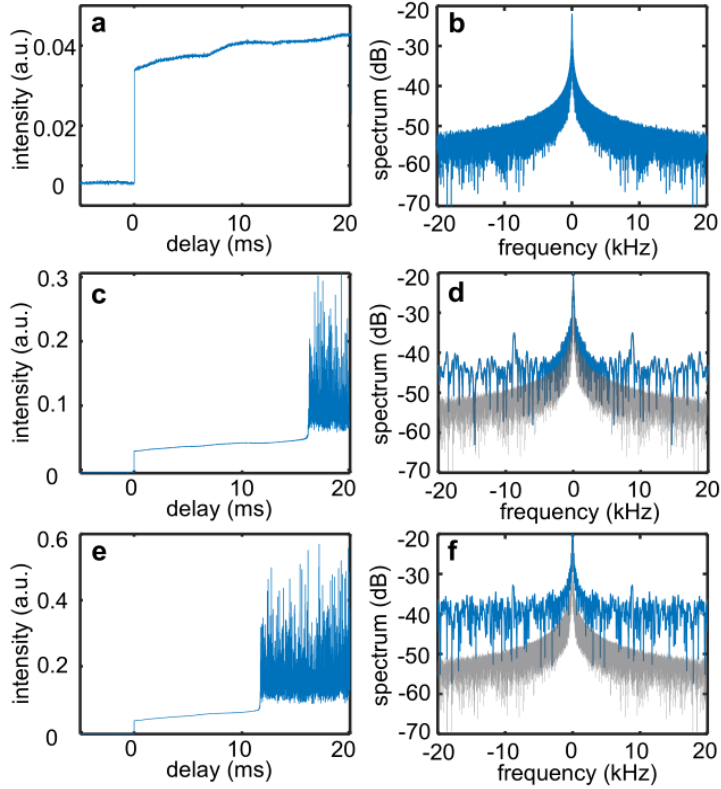

FIG. S6. **Self-organization of time crystal with power dependence.** (a) Time evolution of the measured $I(t)$ for an input of $P_{in}$ = 5 mW. (b) The spectrum corresponds to (a). (c) Measured $I(t)$ for $P_{in}$ = 6 mW. (d) The spectrum corresponds to (c) with the spectrum of $P_{in}$ = 5 mW in the background. (e) Measured $I(t)$ for $P_{in}$ = 7 mW. (f) The spectrum corresponds to (e) with the spectrum of $P_{in}$ = 5 mW in the background. The pump laser frequency is $f_l$ = 1.17 GHz. The time crystal frequency remains 8.7 kHz.

As shown in the main text that the phase transitions of the inherent time crystal depend on the pump power. For different laser frequency $f_l$, i.e., placing the laser frequency in different positions of the inhomogeneous absorption line of the Er:Y$_2$SiO$_5$, different aspects of such phase-transition processes can be unfolded. Here we show more data of the phase transitions at $f_l = 1.17 \text{GHz}$.

When the pump power is low $P_{in} = 5\text{mW}$, there is no oscillating signal in Fig. S6a and S6b. If the pump laser is increased to 6.0mW, we can see that $I(t)$ becomes dynamically unstable, indicating that the time translation symmetry is spontaneously broken by the erbium dipole-dipole interactions, as shown in Fig. S6c. Its corresponding Fourier spectrum shows a periodic time pattern of 8.7kHz, as shown in Fig. S6d, indicating the forming of temporal order. Further increasing the pump power to 7.0mW, the 8.7kHz peak remains, and the changes in its amplitude and width are not obvious. However, oscillations at other frequencies become much more significant, as shown in Fig. S6f. Specifically, the spectral intensities at non-zero frequencies (apart from 8.7kHz) are approximately -45dB in Fig. S6d and -40dB in Fig. S6f. The 8.7kHz peak becomes less obvious as a result of the growing amplitudes of other frequencies. This may finally cause the time crystal to be unidentified. Note that such a transition is different from that of Fig. 3a in the main text, where the broadening of the 8.7kHz peak at strong $P_{\text{in}}$ is caused by the high-order effects rather than growing background noise.

## Supplementary Note 8. Intrinsic optical instability without time crystalline order.

The measured $I(t)$ for $P_{\text{in}} = 8$mW and $f_l = 0.50$GHz corresponding to Fig. 3a in the manuscript is shown in Fig. S7. After the laser is switched on at $t = 0$ms, the system takes approximately 14ms to reach a phase of intrinsic optical instability. However, there is no temporal periodicity in $I(t)$ under this circumstance, as evidenced by the lack of peak in its spectrum measured in the long-time limit, which is shown as the orange line in Fig. 3A in the main text.

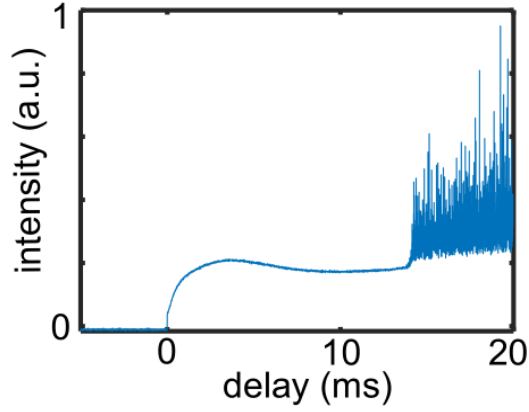

FIG. S7. **Intrinsic optical instability without time crystalline order**. Measured $I(t)$ of $P_{in}$ = 8 mW and $f_l$ = 0.5 GHz. The laser field is switched on at $t = 0$ ms.

## Supplementary Note 9. Data analysis using cross-correlation function

Here we present the method to obtain the phase information of $I(t)$, as shown in Fig. 3C of the main text. Here we consider a periodic function with both amplitude noise and phase noise such that

$$f(t) = [1 + a(t)] \cdot cos[\omega_0 t + \phi(t)], \tag{29}$$

where $\omega_0$ is the time crystal frequency, $a(t)$ is the amplitude noise, and $\phi(t)$ is the phase noise. The function $f(t)$ is used to represent the measured $I(t)$ in experiments. It is worth highlighting here the frequency characteristics of the different parts in $f(t)$:

- The time crystal frequency $\omega_0$ = 8.7 kHz, which is obtained from the autocorrelation and the Fourier spectra of $I(t)$, as shown in Fig. 2 in the main text.

- The intrinsic instability $a(t)$ is considered as a rapidly-varying function of time. The typical frequency of $a(t)$ is on the order of tens of MHz, as detailed in the literature.

- The phase noise $\phi(t)$ is a slowly-varying function of time. According to the ~ 100 Hz linewidth measured in the Fig. 3c in the main text, the corresponding time scale is on the order of milliseconds.

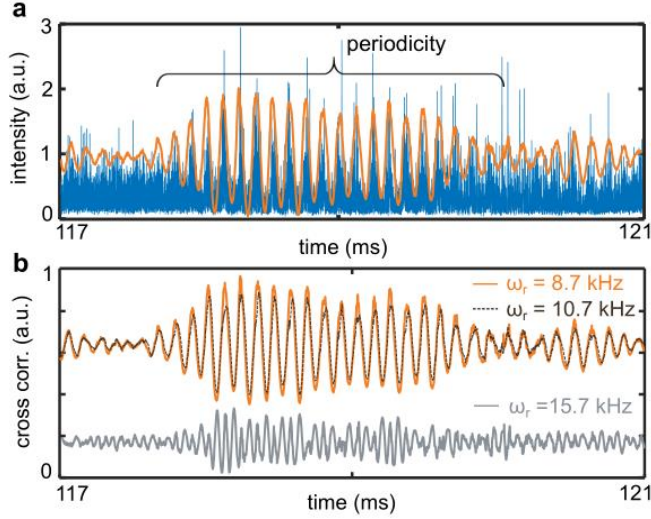

FIG. S8. **Validity of the analysis method.** (a) Comparison of the measured $I(t)$ and the cross-correlation of $F(\tau)$. Blue, the $I(t)$ from 117 ms to 121 ms. Orange, the cross-correlation $F(\tau) = \langle I(t)r(t-\tau)\rangle$ with a delay $\tau$ varying from 117 ms to 121 ms. $\omega_r = 8.7$ kHz and the integration time $T = 0.125$ ms. (b) Weak dependence of $F(\tau)$ on the reference frequency $\omega_r$. Orange solid, $F(\tau)$ is calculated with a reference frequency of $\omega_r = 8.7$ kHz, which is identical to the orange curve in a. Black dashed, $F(\tau)$ calculated with $\omega_r = 10.7$ kHz. Gray solid, $F(\tau)$ calculated with $\omega_r = 15.7$ kHz. The $\omega_r = 15.7$ kHz curve is vertically shifted for clarity. The amplitudes of the three $F(\tau)$ are in scale.

Based on the different time scales involved in $f(t)$, we use an integration method to extract the period and the phase information hidden in the noisy $f(t)$. We first construct a reference pulse function with a frequency of $\omega_r$:

$$r(t) = cos(\omega_r t) \cdot rect(t), \quad \text{with} \quad rect(t) = \begin{cases} 0, & \text{if} \quad t < 0 \\ 1, & \text{if} \quad 0 \leq t \leq T \\ 0, & \text{if} \quad t > T \end{cases} \quad (30)$$

The $rect(t)$ is a rectangle function with an open duration of $T$, and $T$ represents the integration time of our method. The cross-correlation function between $f(t)$ and $r(t)$ is defined as

$$\begin{aligned} F(\tau) &= \int_{-\infty}^{+\infty} f(t)r(t-\tau)dt \\ &= \int_{\tau}^{\tau+T} [1+a(t)] \cdot cos[\omega_0 t + \phi(t)]cos[(\omega_r(t-\tau)]dt \\ &= \int_{\tau}^{\tau+T} \frac{1+a(t)}{2} cos[(\omega_0 + \omega_r)t - \omega_r\tau + \phi(t)]dt \\ &\quad + \int_{\tau}^{\tau+T} \frac{1+a(t)}{2} cos[(\omega_0 - \omega_r)t + \omega_r\tau + \phi(t)]dt. \end{aligned} \quad (31)$$

Typically, we choose $T \sim 0.1$ ms. As aforementioned, $a(t)$ varies rapidly at frequencies $\mathcal{O}(10\text{MHz})$. Thus any terms involved the fast oscillating $a(t)$ is averaged to near zero when calculating the integral. The same conclusion also holds for the term with $\cos(\omega_0 + \omega_r)t$. Using these two approximations, we then have

$$F(\tau) \approx \int_\tau^{\tau+T} \frac{1}{2}\cos[(\omega_0 - \omega_r)t + \omega_r\tau + \phi(t)]dt. \tag{32}$$

Since $\phi(t)$ is a slowly-varying function of time on the order of 1ms, and the integration time $T \sim 0.1\,\text{ms}$, we can replace $\phi(t)$ with $\phi(\tau)$ in the above equation such that

$$\begin{aligned} F(\tau) &\approx \int_\tau^{\tau+T} \frac{1}{2}\cos[(\omega_0 - \omega_r)t + \omega_r\tau + \phi(\tau)]dt \\ &= \frac{1}{\omega_0 - \omega_r}\sin[(\omega_0 - \omega_r)T/2] \cdot \cos[\omega_0\tau + \omega_0 T/2 + \phi(\tau)]. \end{aligned} \tag{33}$$

At the limit that $(\omega_0 - \omega_r)T \to 0$, we obtain that

$$F(\tau) \approx \frac{T}{2} \cdot \cos[\omega_0\tau + \omega_0 T/2 + \phi(\tau)] \tag{34}$$

Neglecting a constant phase of $\omega_0 T/2$, we finally have

$$F(\tau) \propto \cos[\omega_0\tau + \phi(\tau)] \tag{35}$$

Comparing Eq (35) and Eq (29), we can reproduce the period and the phase information of $f(t)$ by choosing different $\tau$ (with a time resolution determined by $T$). To confirm the method's validity, we calculate the cross-correlation between the $I(t)$ from 117ms to 121ms in Fig. S8a and a reference function with $\omega_r = 8.7\text{kHz}$ and $T = 0.125\text{ms}$. The result is plotted as an orange line on top of Fig. S8a. It is obvious that the cross-correlation $F(\tau)$ reproduces the period and the phase information of $I(t)$ from a background of high frequency noise.

Note also that the analysis method here is not sensitive to the reference frequency $\omega_r$. This property can be seen from Eq. (34), which does not depend on $\omega_r$. That is to say, one does not have to precisely pre-measure $\omega_0$ and set $\omega_r = \omega_0$ during the data analysis. As long as $(\omega_r - \omega_0)T \to 0$, Eq. (34) holds. We present $F(\tau)$ of three different $\omega_r$ in Fig. S8b. While the result of $\omega_r = 8.7\text{kHz}$ indicates a periodic oscillation of $I(t)$, the period and the phase information of $I(t)$ can also be revealed, almost identically, by the result of $\omega_r = 10.7\text{kHz}$. However, if $\omega_r$ is further increased to $15.7\text{kHz}$, the $F(\tau)$ becomes improper and is no longer useful in extracting the phase information, as shown by the gray curve in Fig. S8b.

## Supplementary Note 10. Periodicity at different delays

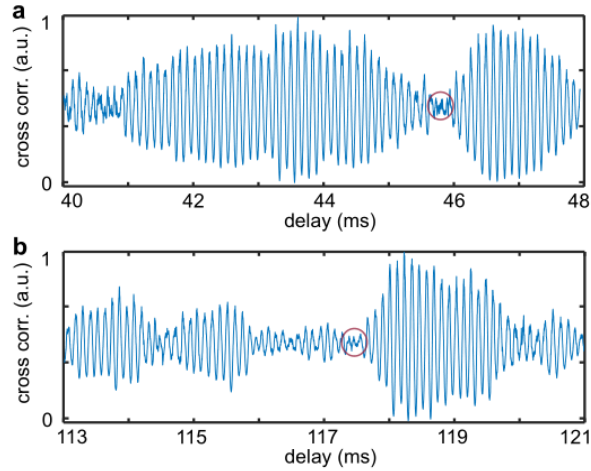

FIG. S9. **The correlation function $F(\tau)$ at different delay times.** (a) $F(\tau)$ of delay $\tau$ from 40-48 ms. A time point with phase discontinuity can be identified at approximately 46 ms. (b) $F(\tau)$ of delay $\tau$ from 113-121 ms. A time point with phase discontinuity can be identified at approximately 117.5 ms.

Using $F(\tau)$ detailed in Note 8, one can reveal the periodicity and the phase information of $I(I)$ more clearly than the raw data $I(t)$ itself. Exemplified in Fig. S9 are the cross correlation $F(\tau)$ for different delay times of the same $I(t)$ in Fig. 3C in the manuscript. It is clear that oscillations with a frequency of 8.7kHz consecutively arise from 40 ms to 121ms, as shown in Fig. S9a, S9b and Fig. 3C in the manuscript. Instead of being a constant in the long-time scale, the phase of oscillations is a slowly-varying function of time, showing phase discontinuities at random time positions, as marked by the red circles in Fig. S9a and S9b. As discussed in the manuscript, such phase discontinuities result from the phase noise of our fiber laser.

The phase change consistently occurs when the 8.7 kHz signal slowly envelopes to a small amplitude (it never occurs when the 8.7 kHz oscillations are of large amplitude), indicating a continuous phase change rather than a sudden jump. However, characterizing the laser's phase noise at 100 Hz scales is challenging, and we lack the necessary equipment for a conclusive determination of its nature.

## Supplementary Note 11. Phase discontinuities in the theoretical model

To understand the effect of the phase noise in driving to the inherent time crystal, we have calculated the response of $\rho(t)$ for a laser pump with phase shift at some specific moments. The results are shown in Fig. S10. We can see that the system takes approximately 4ms to self-organize to temporal order, as shown by the left inset of Fig. S10. At $t = 5.6$ms and $t = 11.2$ms, phase shifts of $\pi$ are imposed on the optical driving field, as marked by the red dashed lines in Fig. S10. These phase perturbations break the balance of the system. The $\rho(t)$ thus losses its temporal order and takes approximately 4ms to self-organize to another one, as shown by the right inset of of Fig. S10. The period of this new temporal order is the same as the previous one. Still, there is a phase discontinuity between them. The calculated results of imposed phase discontinuities in the driving field agree well with our experiment: although the oscillating $I(t)$ persists, the phase noise of the laser breaks $I(t)$ into segments on the order of milliseconds.

## Supplementary Note 12. Differences between self-pulsing and time crystal

It is well-known that self-pulsing effect can occur in erbium-doped fiber lasers , in which the laser output power fluctuates in a periodic or quasi-periodic manner without any external modulation. The fluctuations in the output power are generally caused by the competition between the gain of the laser medium and the losses due to the cavity's various components. As a result, the net gain inside the cavity is periodically modulated and generate fluctuations in the laser output. Although our time crystal and self-pulsing both exhibit periodic modulations in their outputs, they are fundamentally different and have different underlying physical mechanisms. The most important differences are:

1. **The experimental setups are different**. Optical cavity is a necessary for the well-know self-pulsing effects in erbium-doped fibre lasers. Our experimental setup differs significantly. One end of our sample is antireflection coated with a reflectivity of less than 0.8%, preventing the formation of a similar cavity. In addition, our sample is of 12mm long. In contrast, the fibre laser systems that manifest self-pulsing effect typically have a length on the order of meters, that is, two-orders-of-magnitude difference. Thus, the achievable optical gain in our crystal, given that the erbium concentrations are similar, is much less than that in fibre laser systems due to these factors. These differences distinguish our experimental setup from traditional fiber laser systems and should be considered in any comparisons or analyses.

2. **The experimental pumping pumping conditions different.** Self-pulsing behaviour in 1.5$\mu$m erbium-doped fiber lasers are observed with pumping at shorter wavelength, such as 514 nm, 810nm or 980nm. The 1.5 $\mu$m erbium laser which are self-pulsing when pumped at 980 nm becomes stable when pumped at 1490 nm or longer. In fact, it has been demonstrated that by adding an auxiliary 1.5$\mu$m pump with only 3% of the lasing power, the self-pulsation in the system can be significantly suppressed. Our results were observed with erbium ions pumped by 1.5 $\mu$m laser, which can effectively eliminate the self-pulsing effect.

3. **The experimental phenomena are different.** In self-pulsing experiments, increasing the pumping power can drive the laser output from a stable to a periodically modulated regime, and further increase can lead to chaotic behavior. However, in our experiments, increasing the pump power first results in dynamically unstable output with frequency response up to

~50MHz. With additional pump power, a periodic oscillation of 8.7,kHz emerges. Beyond that, higher pump power brings the system to another unstable phase. Moreover, the 8.7 kHz in our experiment is unaffected by the increase of the pump power. This observation is in contrast to the case of self-pulsing, whose frequency typically depends on the pump power. As a result, the behavior of our system under increased pumping power differs significantly from that of self-pulsing systems (see Supplementary Note 6, SI).

4. **The physical origins are different.** The self-pulsing effect is typically the result of the dymamic interplay between the optical gain and the absorption or losses in laser systems. Under these circumstances, the net optical gain can be temporally modulated, leading to the switching on-and-off of the laser output. However, our experiment does not involve a cavity or optical net gain, and cannot be explained as the switching on-and-off of the cavity output. Additionally, the up to ~50MHz frequency response observed in our time crystal phase exceeds the energy transfer rate between the erbium ions in the self-pulsing model. Therefore, the temporal order observed in our experiments likely arises from another mechanism instead of the well-known self-pulsing effect.

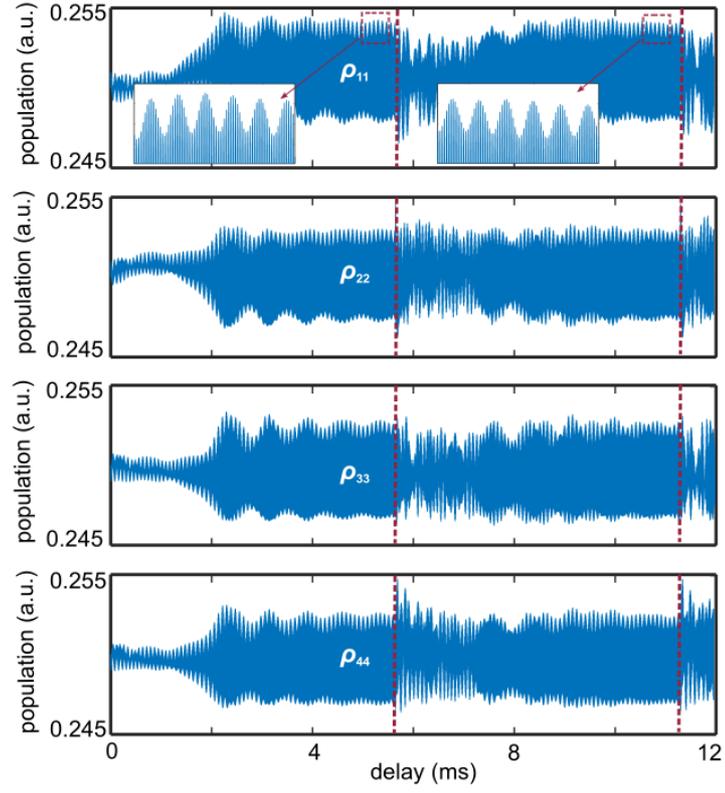

FIG. S10. **Calculation of $\rho_{nn}(t)$ with phase noise in the driving field.** The dynamic behaviors of the populations in the four levels as noted. $\rho_{nn}(t)$ shows persistent oscillations in the long time limit. It takes approximately 4 ms to form an inherent time crystal phase. At $t = 5.6$ ms and $t = 11.2$ ms, as marked by the red dashed line, phase shifts of $\pi$ are imposed to the optical driving field.